\documentclass[structabstract]{raa}
\usepackage{graphicx,times}
\usepackage{natbib,amssymb}
\usepackage{epsfig}
\usepackage{float}
\usepackage{color}
\usepackage[figuresright]{rotating}
\providecommand{\bm}[1]{\mbox{\boldmath$#1$\unboldmath}}
\begin{document}

\title{$R$-band host galaxy contamination of TeV $\gamma$-ray blazar Mrk 501: effects of the aperture size and seeing}

 \volnopage{ {\bf 20xx} Vol.\ {\bf X} No. {\bf XX}, 000--000}
   \setcounter{page}{1}

\author{Hai-Cheng Feng\inst{1,2,3}, Hong-Tao Liu\inst{1,3,4}$^{\bigstar}$\footnotetext{\small $^{\bigstar}$ Corresponding authors:
H. T. Liu, e-mail: htliu@ynao.ac.cn; Ying-He Zhao: zhaoyinghe@ynao.ac.cn.}, Ying-He Zhao\inst{1,3,4}$^{\bigstar}$, Jin-Ming Bai\inst{1,3,4},
 Fang Wang\inst{5}, Xu-Liang Fan\inst{6}}

   \institute{Yunnan Observatories, Chinese Academy of Sciences, 396 Yangfangwang, Guandu District, Kunming, 650216, P. R. China\\
   \and
    University of Chinese Academy of Sciences, Beijing 100049, P. R. China\\
    \and
    Key Laboratory for the Structure and Evolution of Celestial Objects, Chinese Academy of Sciences, 396 Yangfangwang,
    Guandu District, Kunming, 650216, P. R. China\\
    \and
    Center for Astronomical Mega-Science, Chinese Academy of Sciences, 20A Datun Road, Chaoyang District, Beijing, 100012, P. R. China\\
    \and
    School of Physics and Space Science, China West Normal University, Nanchong, 637009, P. R. China\\
    \and
    School of Physics, Huazhong University of Science and Technology, Wuhan 430074, P. R. China\\
}
 \date{Received ; Accepted}

 \abstract{We simulated the $R$-band contribution of the host galaxy of TeV $\gamma$-ray BL Lac object Mrk 501 in different aperture
 sizes and seeing conditions . The intensive observations were run with the 1.02 m optical telescope at Yunnan Observatories from 2010
 May 15 to 18. Based on the host subtraction data presented in Nilsson et al. (2007), the subtraction of host galaxy contamination results
 in significant seeing-brightness correlations. These correlations would lead to illusive large amplitude variations at short timescales,
 which will mask the intrinsic micro variability, thus gives rise to difficulty in detecting the intrinsic micro variability. Both aperture size
 and seeing condition influence the flux measurements, but aperture size impact the result more significantly. Based on the parameters
 of elliptical galaxy provided in Nilsson et al. (1999), we simulated the host contributions of Mrk 501 in the different aperture sizes and
 seeing conditions. Our simulation data of the host galaxy obviously weaken these significant seeing-brightness correlations for the
 host-subtracted brightness of Mrk 501, and can help us discover the intrinsic short timescale micro variability. The pure nuclear flux is
 $\sim$ 8.0 $\rm{mJy}$ in \emph{R} band, i.e., AGN has a magnitude of $R\sim13^{m}_{\cdot}96$.
  }

   \authorrunning{H.-C. Feng et al. }            %author_head in even pages
   \titlerunning{$R$-band host galaxy contamination of Mrk 501}  % title_head in odd pages

   \maketitle

%________________________________________________ sections below
%
\section{Introduction}           %% first-level sections will be auto-capitalized
\label{sect:intro}

   Blazars are an extreme subclass of active galactic nuclei (AGNs), including BL Lacertae (BL Lac) objects and flat spectrum
   radio quasars (FSRQs) (e.g., Angel \& Angel 1980; Urry \& Padovani 1995; Fossati et tal. 1998; B\"{o}ttcher \& Dermer 2002;
   Maraschi \& Tavecchio 2003). They are characterized by rapid and strong variability over the whole electromagnetic spectrum,
   high and variable polarization from the optical to radio bands, and prominent non-thermal emission at all wavelengths. In general,
   these extreme properties can be generated from a relativistic jet with a viewing angle less than $10^{\circ}$ (e.g., Blandford \&
   K\"{o}nigl 1979; Urry \& Padovani 1995). The broadband spectral energy distributions (SEDs) of blazars usually exhibit a double
   peak profile. The first component extends from infrared to ultraviolet or soft X-ray, and the second is located in the GeV/TeV
   gamma-ray bands (e.g., Ghisellini et al. 1998; Abdo et al. 2010). The first peak is generally believed to be the synchrotron radiation
   of relativistic electrons in the jet. The second peak is attributed to the inverse-Compton scattering of the same electron
   population that produces the synchrotron radiation (e.g., Dermer \& Schlickeiser 1993; B\"{o}ttcher 2007; Neronov et al. 2012).

   Due to the property of strong variability of BL Lac object, the photometric technique is widely used to investigate the structure,
   radiation mechanism, dynamics, and the masses of the central supermassive black holes (e.g., Ciprini et al. 2003, 2007; Gupta
   et al. 2008a; Liu \& Bai 2015; Dai et al. 2015). However, the host galaxies often exhibit strong radiation in the optical to near-infrared
   (NIR) bands. Thus, the contamination from the host galaxies may influence the photometric results, especially for nearby extended
   sources. The photometric aperture is either a dynamic aperture or a fixed aperture. The dynamic aperture could be a few
   times the seeing, and the case of an extended source will result in a significant dependence of the photometric magnitudes on
   the seeing. There is not the dependence on the seeing for a point source at high redshift. The fixed aperture and the dynamic
   aperture could result in similar influences on the photometric results for the extended source due to the seeing (see Feng et al.
   2017). For point sources, the strong host galaxies could dilute the variability amplitudes of AGNs. Besides, the color indices and
   the SEDs of AGNs will be influenced. Since an extended source is resolved, different aperture sizes and seeing conditions
   would introduce large uncertainties in photometry at different epochs.

   However, the host galaxies of nearby BL Lac objects are elliptical galaxies, which are huge (the effective radius $R_{e} \sim$ 10 kpc)
   and luminous ($M_{R} \sim -24^{m}_{\cdot}0$) (e.g., Falomo \& Kotilainen 1999; Urry et al. 1998, 2000; Scarpa et al. 2000; Kotilainen
   \& Falomo 2004; Nilsson et al. 2003, 2007; Hyv\"{o}nen et al. 2007). Even though some BL Lac objects may show signs of interaction
   with companions (e.g., Stickel et al. 1993; Falomo 1996; Heidt et al. 1999; Falomo \& Ulrich 2000), there is no clear evidence in most
   cases that the nuclear activity is triggered by interaction (Nilsson et al. 1999, 2007). For most BL Lac objects, the morphologies of
   host galaxies are indistinguishable from the similar normal elliptical galaxies (Scarpa et al. 2000). Thus, the host galaxies of
   BL Lac objects can be simulated based on the normal elliptical galaxies.

   Mrk 501 is a prototype nearby BL Lac object (redshift $z=0.034$), which has been widely studied over the two decades in the
   entire electromagnetic spectrum (e.g., Stickel et al. 1993; Quinn et al. 1996; Catanese et al. 1997; Samuelson et al. 1998; Xie et al. 1999,
   2001; Konopelko et al. 2003; Gupta et al. 2008b, 2012; Albert et al. 2007; Abdo et al. 2011; Shukla et al. 2015; Ahnen et al. 2017). In the
   high energy regime from X-ray to TeV, Mrk 501 is one of the brightest extragalactic sources  (Abdo et al. 2011). Many studies attempted
   to investigate its properties in the optical bands (e.g., Xie et al. 1999, 2001; Gupta et al. 2008b, 2012; Xiong et al. 2016). Based on the
   host subtraction data presented in Nilsson et al. (2007), widely used in the previous photometric studies, the subtraction of host
   galaxy contamination results in a significant seeing-magnitude correlation for Mrk 501 (Feng et al. 2017). The researches related to
   the variability will need a reasonable subtraction of the host galaxy, which should eliminate (partly) this significant seeing-brightness
   correlation.

   In this paper, we presented observations of Mrk 501 in the $R$ band from 2010 May 15 to 18. In order to obtain the host components
   in the different aperture radii and seeing conditions, we used the two-dimension simulation method to produce the host galaxy. The
   structure of this paper is as follows: Section 2 is the observations and data reduction; Section 3 gives the details of simulations; Section 4
   is conclusions, and discussion is presented in Section 5.

\section{Observations and Data Reduction}

   The observations of Mrk 501 was carried out with the 1.02 m optical telescope at Yunnan Observatories. This telescope
   is a classical Cassegrain telescope located at Kunming, China. An Andor AW436 CCD (2048 pixels $\times$ 2048 pixels)
   camera was equipped at the $f$/13.3 Cassergrain focus of the 1.02 m telescope. The entire field of view of the CCD is
   $\sim 7.3 \times 7.3 \/\ \rm{arcmin}^2$, and each pixel corresponds to 0.21 arcsec in both dimensions. The CCD readout
   noise and gain are 6.33 electrons and 2.0 electrons/ADU, respectively (e.g., Dai et al. 2015; Xiong et al. 2016). We selected
   the standard Johnson broadband filters to carry out the observations in the $R$ band, and 326 valid exposures were
   obtained in 4 nights from 2010 May 15 to 2010 May 18. The exposure time is 150 seconds for each frame. Table 1 presents
   the complete observation log. For each image, the standard stars are always in the same field with the object.
\begin{table}
\begin{center}
\caption[]{Observation log and results of IDV observations of Mrk 501.}
\label{Table1}
 \begin{tabular}{cccc}
  \hline\noalign{\smallskip}
   Date & N & Exposure (s) & $\sigma$(star1-star6)   \\
  \hline\noalign{\smallskip}
   2010 May 15	&	88  &  150	&	0.005		\\
   2010 May 16	&	88	&  150  &	0.007		\\
   2010 May 17	&	80	&  150  &	0.005		\\
   2010 May 18	&	70	&  150  &	0.005		\\
  \noalign{\smallskip}\hline
\end{tabular}
\end{center}
  \tablecomments{0.86\textwidth}{Column 1: observation dates; Column 2: observation numbers;
   Column 3: exposure time; Column 4: standard deviation of the (star 1 - star 6).}
\end{table}

   Because the magnitudes of the standard stars are considered constant, the brightness of the object could be calibrated using the
   standard stars  (e.g., Bai et al. 1998; Fan et al. 2014; Zhang et al. 2004, 2008). There are 6 standard stars, whose magnitudes have
   been measured in other works, in the field. (Villata et al. 1998; Fiorucci \& Tosti 1996). In order to improve the measurement accuracy
   of the object magnitude, the selection of the standard stars should consider both of the position in the field and the brightness.
   Star 1 is the brightest of all nearby standard stars [see Figure 9 in Villata et al, (1998) for numbering], and is very close to the object.
   Thus, we selected star 1 to calculate the object magnitude. However, there are some uncertainties, which may introduce some errors
   to the standard stars, i.e. the relative brightness of the standard stars may change in some images. So another comparison star is
   necessary. Star 6 is the closest to the object, and is used as another standard star. We used the standard deviation of the star 1 and star 6 [$\sigma$(star1-star6)] to characterize the change. The standard deviation of the differential instrumental magnitude of star1-star6 is
   $\sim$ 0.005 (see Table 1).

   All of the observed data was reduced using the standard procedure in the Image Reduction and Analysis Facility (IRAF) software.
   For each night, we took the median of all the bias frames and generated a master bias. Then the master bias was subtracted from
   all the object image frames and flat-field image frames. We used the same method to generate the master flat-field, and then
   the flat-field correction was performed. After the corrections of bias and flat-field, aperture photometry was performed using the
   APPHOT task. Considering the standard stars are point sources, an extraction aperture depending on full width at half maximum
   (FWHM), i.e., a dynamic aperture, was used to obtain the maximum signal-to-noise ratio (S/N) (Howell 1989). We found that the
   best S/N was obtained with the aperture radius of 1.2 FWHM [minimizing $\sigma$(star1-star6)]. For the target, we chose 19 fixed
   aperture radii from 1 arcsec to 10 arcsec to investigate the property of the host galaxy. The epoch, differential magnitude, and FWHM
   of each image are listed in Tables 2-5. Figure 1 shows the relationship between the FWHMs and the magnitudes in different apertures
   for each night, and Figure 2 shows the corresponding relationship of the FWHMs and the fluxes. Figures 1 and 2 indicate that both
   of the FWHM and aperture affect the photometric results. The brightness increases as the aperture increases, and decreases as the
   seeing increases. The increasing aperture will contain more light, and the increasing seeing will scatter more light out of the aperture.
\begin{table}
\begin{center}
\begin{minipage}[]{100mm}
\caption[]{The observed data on 2010 May 15}
\label{Table2}
 \end{minipage}
 \setlength{\tabcolsep}{2.5pt}
\small
\begin{tabular}{cccccccccccccccc}
\hline\noalign{\smallskip}
  MJD  &   &   &   &  & &  Apert &    &    &   &  & &   &   &   &  FWHM  \\ \cline{2-15}

  (day)  &1.0  & 1.5  &2.0   & 2.5 &  3.0  & 3.5   & 4.0  & 4.5   &5.0   &5.5 & 6.0 & 6.5 & ...  & 10.0& (arcsec)  \\

\hline\noalign{\smallskip}
5331.699363  & 4.121  & 6.985  & 9.483  &11.412  &12.958  &14.235  &15.239  &16.179  &16.988  &17.740  &18.423  &19.044  & ...  &22.416  & 1.98  \\
5331.701875  & 4.335  & 7.214  & 9.651  &11.581  &13.102  &14.353  &15.380  &16.299  &17.114  &17.871  &18.542  &19.167  & ...  &22.519  & 1.86  \\
5331.703727  & 4.307  & 7.174  & 9.597  &11.475  &12.970  &14.182  &15.183  &16.105  &16.926  &17.642  &18.287  &18.904  & ...  &22.128  & 1.89  \\
5331.705590  & 4.256  & 7.147  & 9.588  &11.486  &12.970  &14.195  &15.197  &16.105  &16.910  &17.642  &18.287  &18.886  & ...  &22.088  & 1.93  \\
5331.707442  & 4.271  & 7.167  & 9.615  &11.496  &12.994  &14.222  &15.211  &16.105  &16.895  &17.610  &18.271  &18.904  & ...  &22.190  & 1.90  \\
5331.711146  & 4.056  & 6.921  & 9.448  &11.433  &13.006  &14.287  &15.323  &16.254  &17.082  &17.789  &18.474  &19.096  & ...  &22.374  & 2.00  \\
 ...          & ...    & ...    & ...    & ...    & ...    & ...    & ...    & ...    & ...    & ...    & ...    & ...    & ...    & ... & ...   \\
5331.692500  & 3.950  & 6.776  & 9.293  &11.276  &12.840  &14.156  &15.183  &16.149  &16.973  &17.724  &18.406  &19.009  & ...  &22.272  & 2.00 \\
\noalign{\smallskip}\hline
\end{tabular}
\end{center}
 \tablecomments{0.86\textwidth}{This table is available in its entirety in a machine-readable form in the online journal. A portion is shown here for
guidance regarding its form and content. MJD = JD - 2450000. Apert: aperture radius in units of arcsec, presented in columns 2--15. The fluxes are in unit of mJy.}
\end{table}

\begin{table}
\begin{center}
\begin{minipage}[]{100mm}
\caption[]{The observed data on 2010 May 16}
\label{Table3}
 \end{minipage}
 \setlength{\tabcolsep}{2.5pt}
\small
\begin{tabular}{cccccccccccccccc}
\hline\noalign{\smallskip}
  MJD  &   &   &   &  & &  Apert &    &    &   &  & &   &   &   &  FWHM  \\ \cline{2-15}

  (day)  &1.0  & 1.5  &2.0   & 2.5 &  3.0  & 3.5   & 4.0  & 4.5   &5.0   &5.5 & 6.0 & 6.5 & ...  & 10.0& (arcsec)  \\

\hline\noalign{\smallskip}
5332.644491  & 3.160  & 5.719  & 8.169  &10.227  &11.829  &13.139  &14.156  &15.071  &15.869  &16.586  &17.241  &17.855  & ...  &21.152  & 2.63  \\
5332.647292  & 3.349  & 5.994  & 8.483  &10.475  &12.027  &13.272  &14.274  &15.197  &15.972  &16.663  &17.288  &17.871  & ...  &20.881  & 2.38  \\
5332.649294  & 3.383  & 6.050  & 8.514  &10.456  &11.983  &13.211  &14.195  &15.071  &15.840  &16.541  &17.161  &17.740  & ...  &20.785  & 2.40  \\
5332.651157  & 3.446  & 6.123  & 8.633  &10.660  &12.273  &13.557  &14.580  &15.508  &16.329  &17.051  &17.675  &18.271  & ...  &21.308  & 2.27  \\
5332.653009  & 3.501  & 6.202  & 8.681  &10.689  &12.262  &13.519  &14.553  &15.479  &16.284  &17.004  &17.626  &18.237  & ...  &21.328  & 2.28  \\
5332.654861  & 3.282  & 5.907  & 8.405  &10.427  &12.038  &13.346  &14.393  &15.309  &16.105  &16.833  &17.480  &18.070  & ...  &21.210  & 2.45  \\
 ...         & ...    & ...    & ...    & ...    & ...    & ...    & ...    & ...    & ...    & ...    & ...    & ...    & ...  & ...    & ...   \\
5332.829722  & 3.782  & 6.543  & 9.040  &10.999  &12.524  &13.771  &14.769  &15.651  &16.434  &17.146  &17.805  &18.389  & ...  &21.506  & 2.13 \\
\noalign{\smallskip}\hline
\end{tabular}
\end{center}
 \tablecomments{0.86\textwidth}{This table is available in its entirety in a machine-readable form in the online journal. A portion is shown here for
guidance regarding its form and content. MJD = JD - 2450000. Apert: aperture radius in units of arcsec, presented in columns 2--15. The fluxes are in unit of mJy.}
\end{table}

\begin{table}
\begin{center}
\begin{minipage}[]{100mm}
\caption[]{The observed data on 2010 May 17}
\label{Table4}
 \end{minipage}
 \setlength{\tabcolsep}{2.5pt}
\small
\begin{tabular}{cccccccccccccccc}
\hline\noalign{\smallskip}
  MJD  &   &   &   &  & &  Apert &    &    &   &  & &   &   &   &  FWHM  \\ \cline{2-15}

  (day)  &1.0  & 1.5  &2.0   & 2.5 &  3.0  & 3.5   & 4.0  & 4.5   &5.0   &5.5 & 6.0 & 6.5 & ...  & 10.0& (arcsec)  \\

\hline\noalign{\smallskip}
5333.686539  & 4.079  & 6.921  & 9.370  &11.245  &12.734  &13.924  &14.906  &15.782  &16.571  &17.256  &17.888  &18.474  & ...  &21.565  & 2.01  \\
5333.688796  & 4.053  & 6.902  & 9.327  &11.183  &12.617  &13.771  &14.728  &15.565  &16.329  &17.004  &17.610  &18.187  & ...  &21.289  & 2.05  \\
5333.690648  & 3.803  & 6.591  & 9.081  &11.009  &12.524  &13.745  &14.742  &15.637  &16.419  &17.114  &17.773  &18.338  & ...  &21.486  & 2.19  \\
5333.692500  & 3.931  & 6.745  & 9.174  &11.050  &12.501  &13.695  &14.634  &15.494  &16.254  &16.926  &17.529  &18.086  & ...  &21.074  & 2.13  \\
5333.694352  & 3.761  & 6.543  & 9.006  &10.898  &12.341  &13.532  &14.486  &15.337  &16.090  &16.755  &17.352  &17.904  & ...  &20.843  & 2.18  \\
5333.696215  & 4.193  & 7.069  & 9.492  &11.338  &12.757  &13.911  &14.837  &15.680  &16.419  &17.082  &17.707  &18.271  & ...  &21.250  & 1.98  \\
 ...         & ...    & ...    & ...    & ...    & ...    & ...    & ...    & ...    & ...    & ...    & ...    & ...    & ...  & ...    & ...   \\
5333.836632  & 3.645  & 6.382  & 8.883  &10.928  &12.547  &13.860  &14.947  &15.898  &16.755  &17.529  &18.203  &18.834  & ...  &22.128  & 2.17 \\
\noalign{\smallskip}\hline
\end{tabular}
\end{center}
 \tablecomments{0.86\textwidth}{This table is available in its entirety in a machine-readable form in the online journal. A portion is shown here for
guidance regarding its form and content. MJD = JD - 2450000. Apert: aperture radius in units of arcsec, presented in columns 2--15. The fluxes are in unit of mJy.}
\end{table}

\begin{table}
\begin{center}
\begin{minipage}[]{100mm}
\caption[]{The observed data on 2010 May 18}
\label{Table5}
 \end{minipage}
 \setlength{\tabcolsep}{2.5pt}
\small
\begin{tabular}{cccccccccccccccc}
\hline\noalign{\smallskip}
  MJD  &   &   &   &  & &  Apert &  &    &   &  & &   &   &   &  FWHM  \\ \cline{2-15}

  (day)  &1.0  & 1.5  &2.0   & 2.5 &  3.0  & 3.5   & 4.0  & 4.5   &5.0   &5.5 & 6.0 & 6.5 & ...  & 10.0& (arcsec)  \\

\hline\noalign{\smallskip}
5334.700336  & 3.717  & 6.489  & 8.949  &10.858  &12.387  &13.607  &14.593  &15.494  &16.299  &17.020  &17.691  &18.304  & ...  &21.525  & 2.19  \\
5334.702801  & 3.530  & 6.225  & 8.689  &10.699  &12.296  &13.607  &14.674  &15.623  &16.465  &17.225  &17.904  &18.508  & ...  &21.765  & 2.27  \\
5334.704664  & 4.019  & 6.838  & 9.301  &11.255  &12.804  &14.065  &15.099  &16.031  &16.864  &17.626  &18.287  &18.921  & ...  &22.251  & 1.99  \\
5334.706516  & 3.782  & 6.561  & 9.023  &10.958  &12.478  &13.720  &14.715  &15.608  &16.404  &17.098  &17.756  &18.355  & ...  &21.525  & 2.12  \\
5334.708368  & 3.772  & 6.543  & 9.006  &10.969  &12.536  &13.796  &14.823  &15.753  &16.556  &17.288  &17.970  &18.576  & ...  &21.805  & 2.14  \\
5334.710231  & 3.619  & 6.323  & 8.801  &10.788  &12.409  &13.733  &14.796  &15.767  &16.632  &17.384  &18.070  &18.713  & ...  &22.006  & 2.17  \\
 ...          & ...    & ...    & ...    & ...    & ...    & ...    & ...    & ...    & ...    & ...    & ...    & ...    & ...    & ... & ...   \\
5334.833229  & 3.652  & 6.400  & 8.883  &10.828  &12.352  &13.594  &14.607  &15.522  &16.329  &17.051  &17.691  &18.287  & ...  &21.545  & 2.19 \\
\noalign{\smallskip}\hline
\end{tabular}
\end{center}
 \tablecomments{0.86\textwidth}{This table is available in its entirety in a machine-readable form in the online journal. A portion is shown here for
guidance regarding its form and content. MJD = JD - 2450000. Apert: aperture radius in units of arcsec, presented in columns 2--15. The fluxes are in unit of mJy.}
\end{table}
\begin{figure}
 \begin{center}
  \includegraphics[angle=0,scale=0.4]{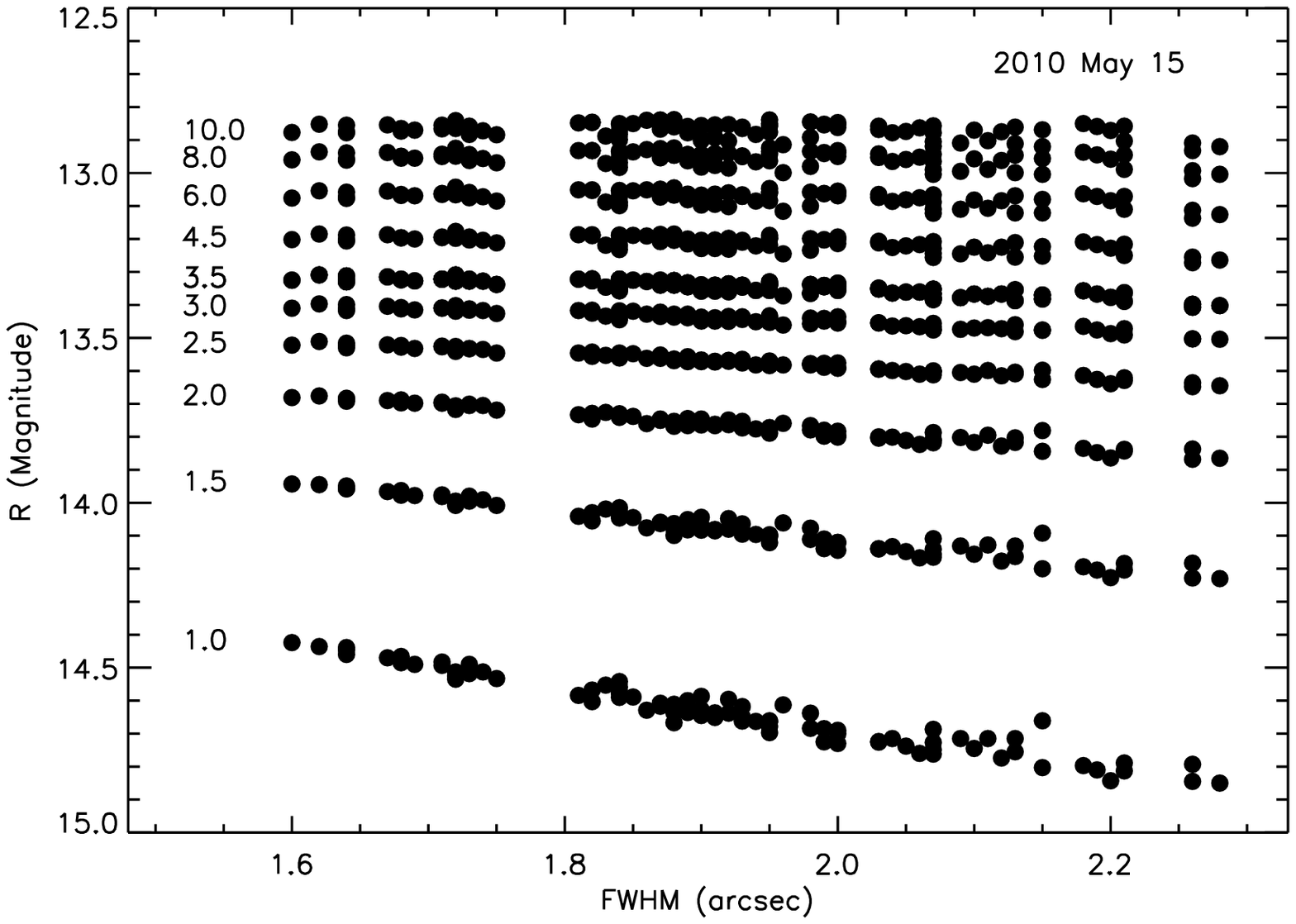}
  \includegraphics[angle=0,scale=0.4]{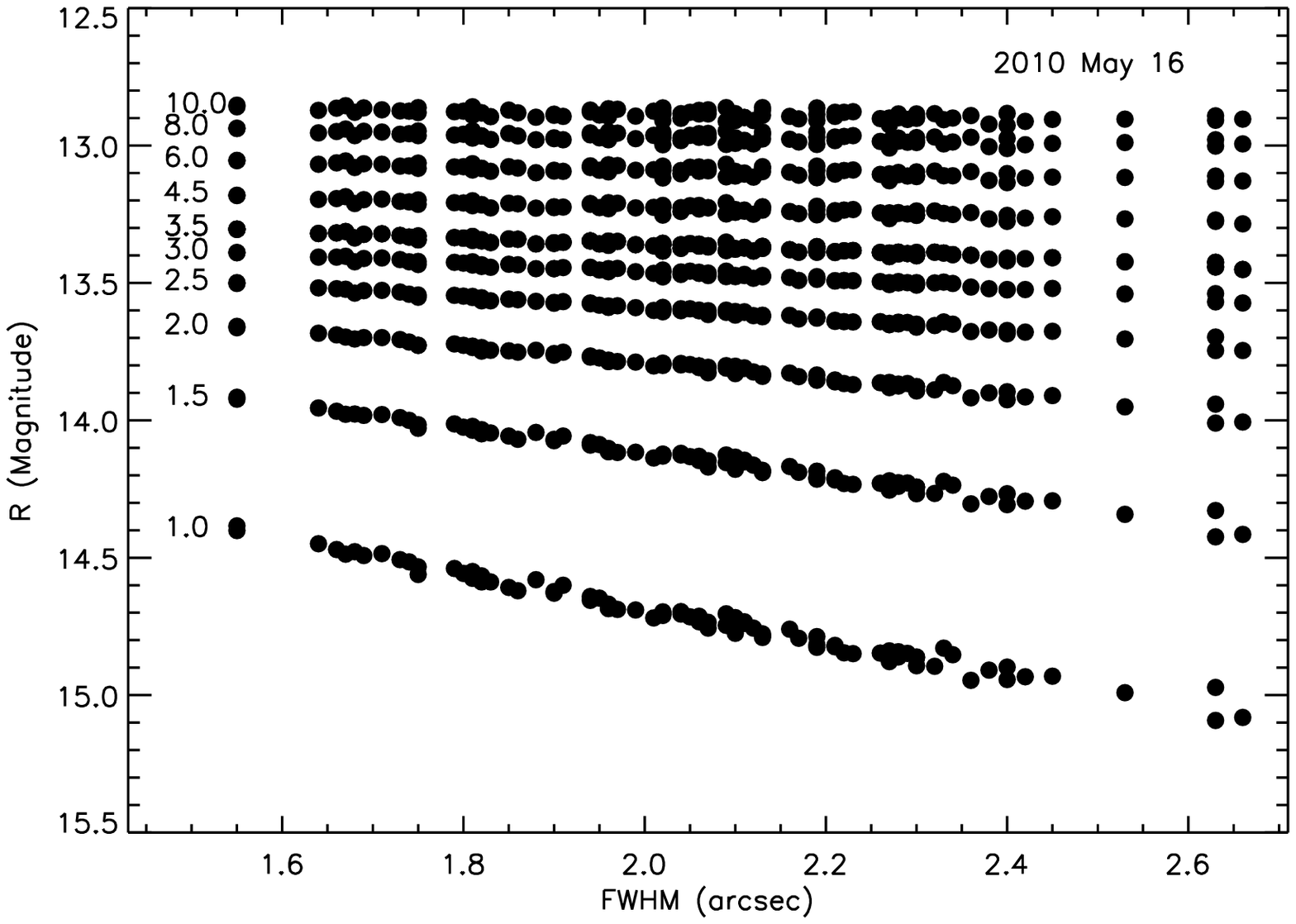}
  \includegraphics[angle=0,scale=0.4]{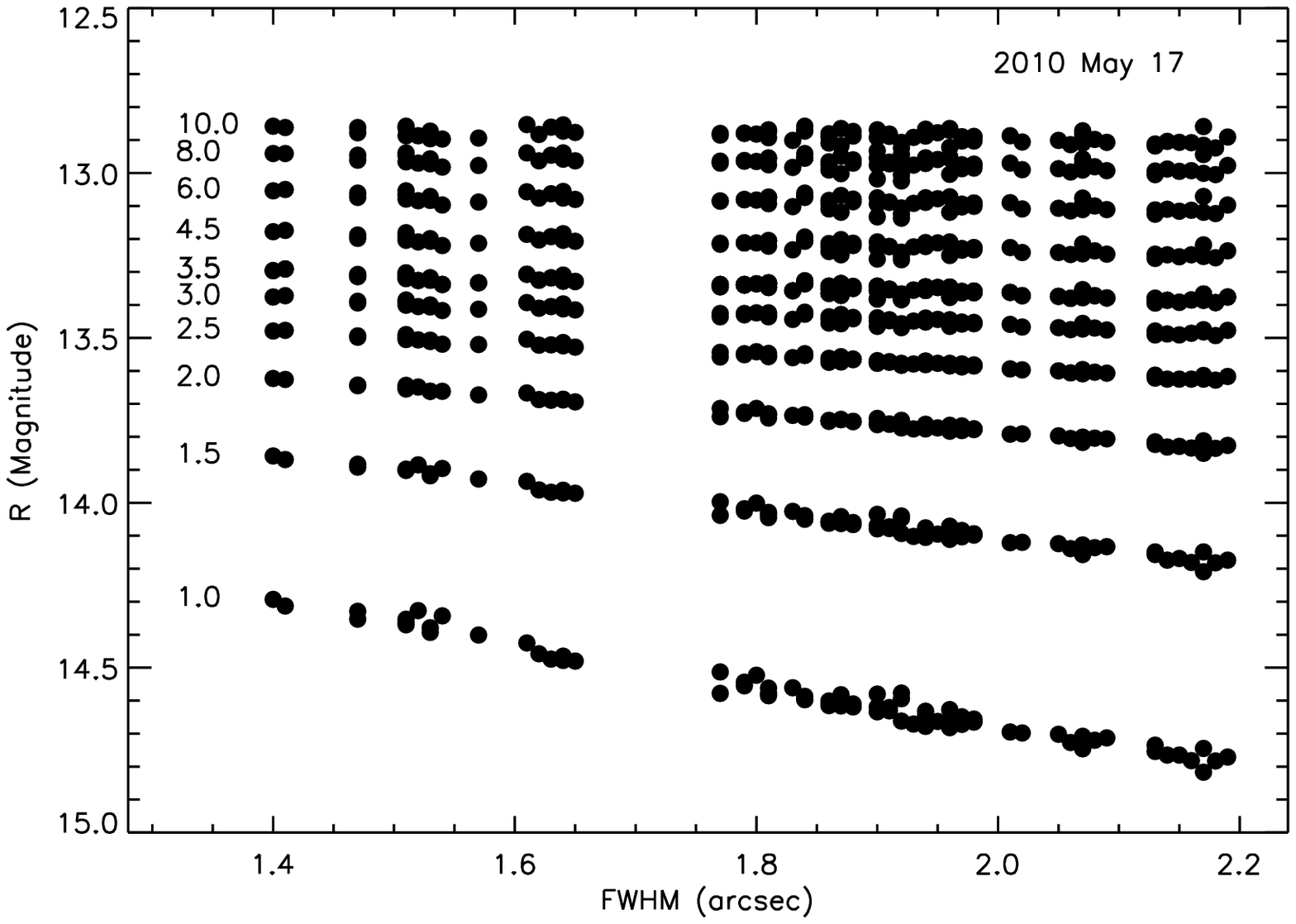}
  \includegraphics[angle=0,scale=0.4]{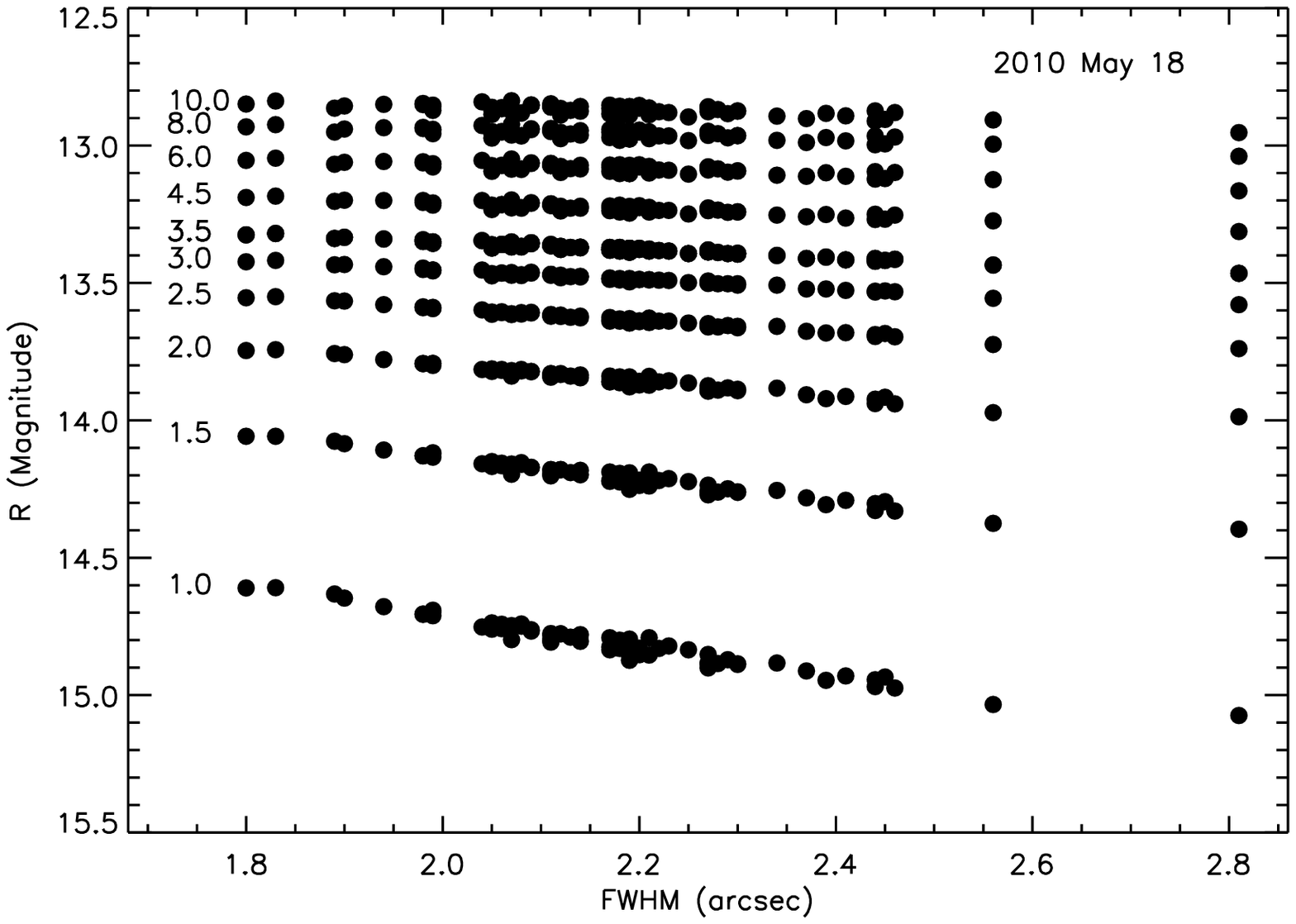}
 \end{center}
 \caption{The relationships between FWHM and magnitude for different photometric aperture radii in our observations.}
  \label{fig1}
\end{figure}
\begin{figure}
 \begin{center}
  \includegraphics[angle=0,scale=0.4]{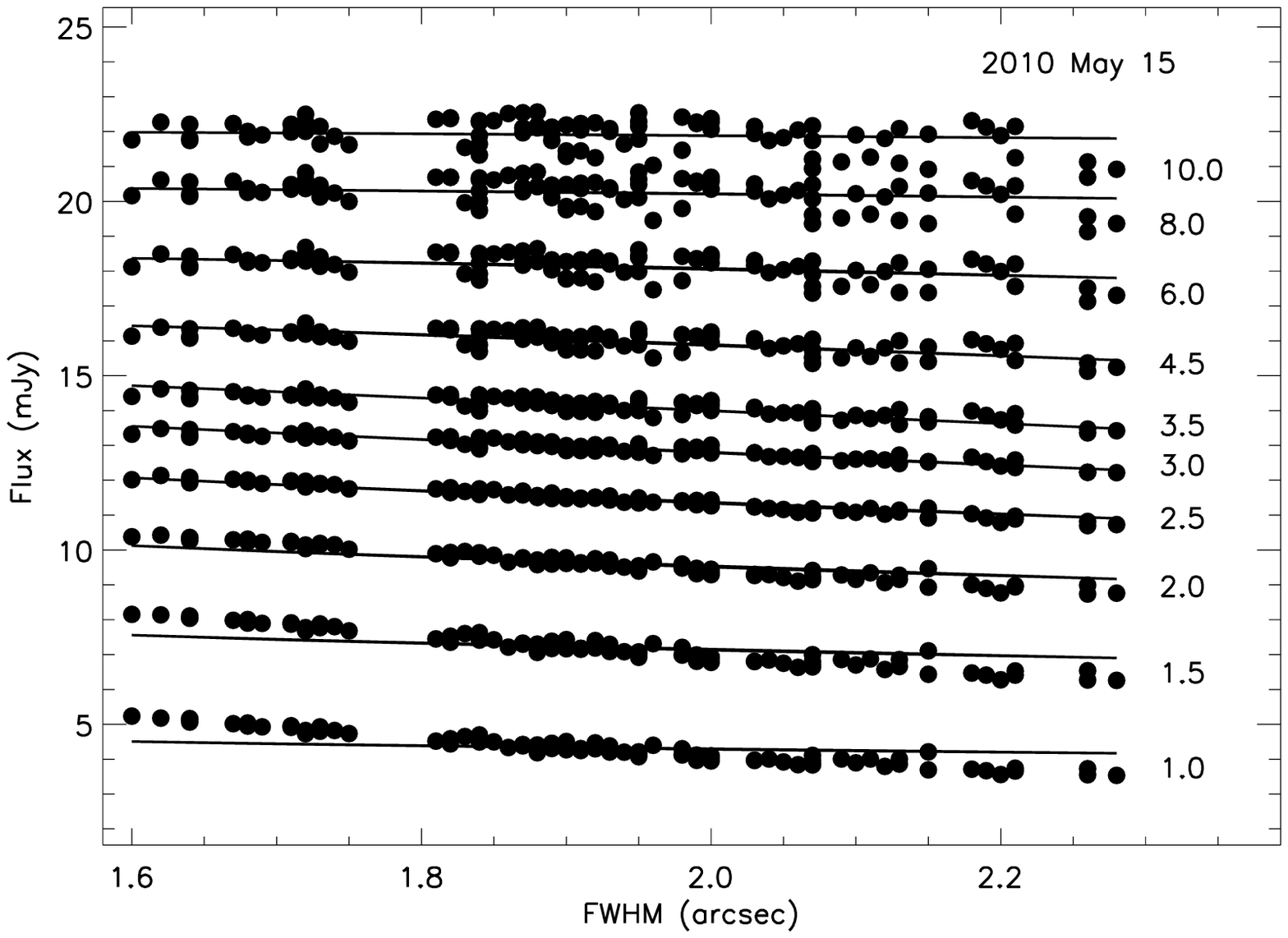}
  \includegraphics[angle=0,scale=0.4]{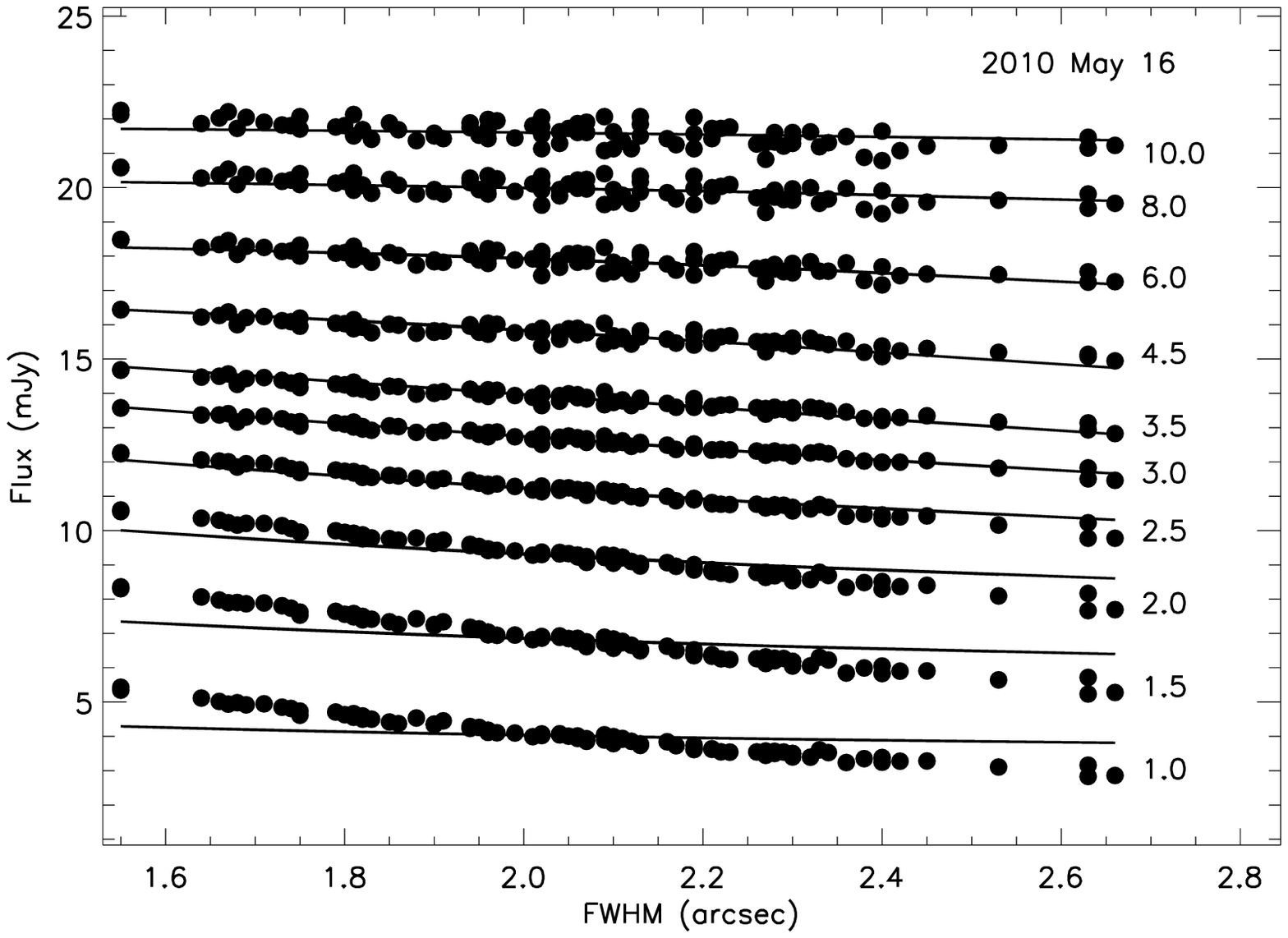}
  \includegraphics[angle=0,scale=0.4]{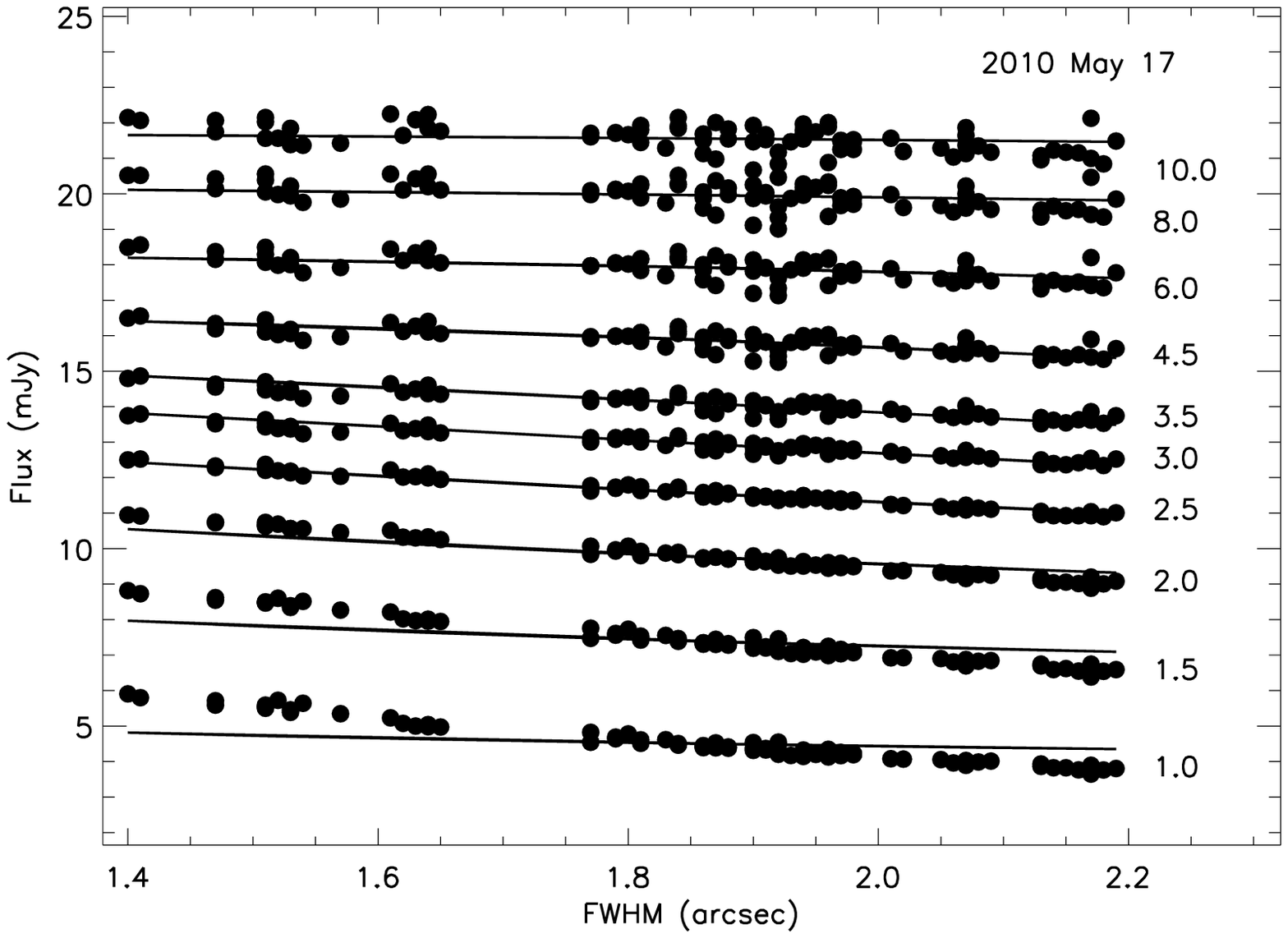}
  \includegraphics[angle=0,scale=0.4]{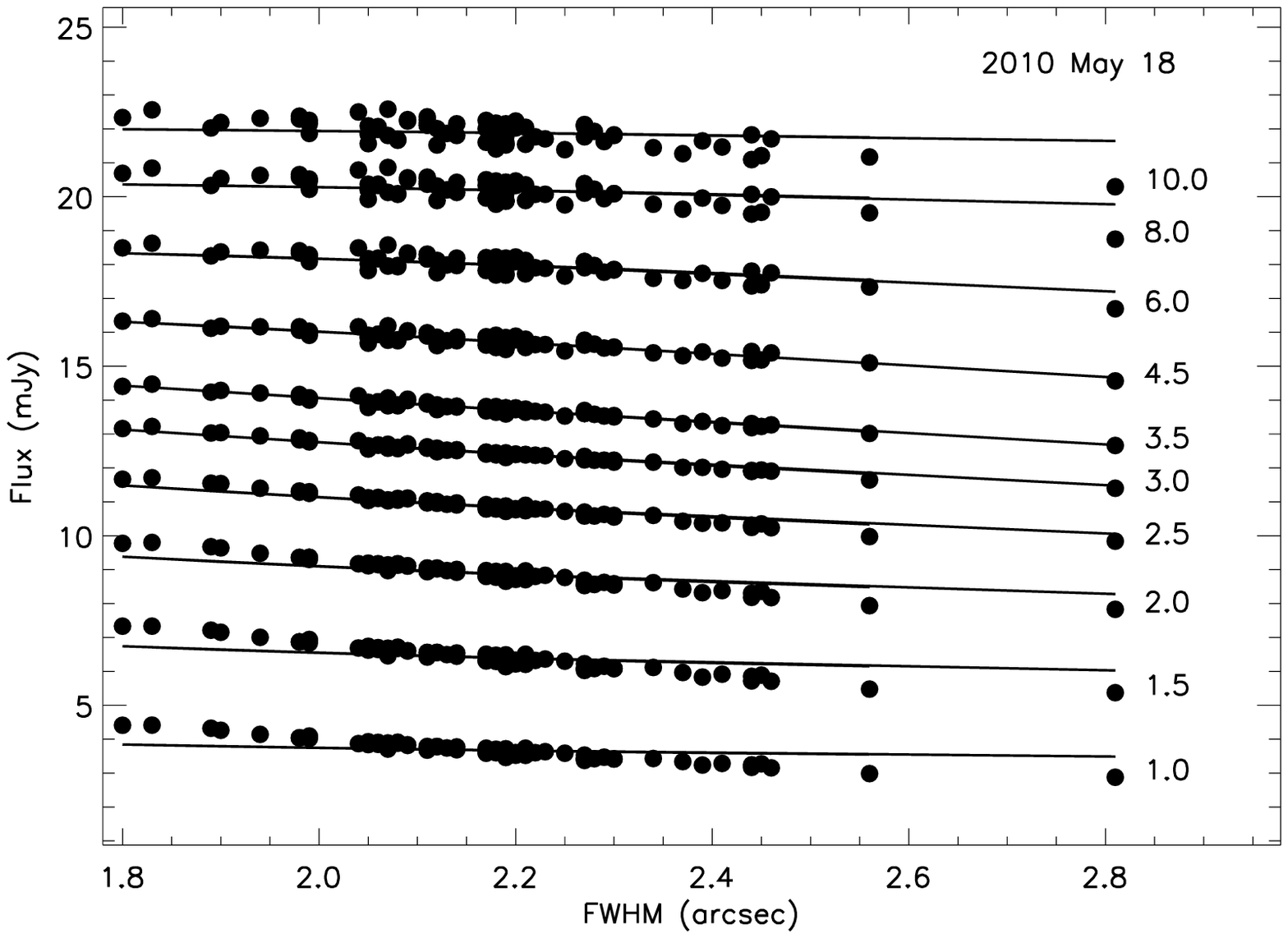}
 \end{center}
 \caption{The relationships between FWHM and flux for different photometric aperture radii in our observations.
  The solid lines are the simulations vertically moved by the average differences between the original simulations
  and the corresponding observational data (the solid circles).}
  \label{fig2}
\end{figure}

\section{Simulations of Host Galaxy}

   The host galaxy of Mrk 501 is an elliptical galaxy (e.g., Nilsson et al. 1999, 2003; Hyv\"{o}nen et al. 2007). Thus, we simulated
   the host galaxy using a two-dimensional model, which assumes the surface brightness $I(r)$ follows the S\'{e}rsic law
   $\sim r^{\rm{\beta}}$ (S\'{e}rsic 1968; Caon et al. 1993; Nilsson et al. 1999). The formula of $I(r)$  is
   \begin{equation}
   I(r)=I(r_{\rm{e}}) dex \left\{ -b_{\beta} \left[ \left( \frac{r}{r_{\rm{e}}} \right)^{-\beta}-1 \right] \right\},
   \end{equation}
   where $\beta$ is the shape parameter, $r_{\rm{e}}$ is the effective radius (containing half of the total luminosity), a $\beta$-dependent
    constant $b_{\rm{\beta}}$ is defined as
   \begin{equation}
   b_{\rm{\beta}}=\frac{0.868}{\beta}-0.142,
   \end{equation}
   and
   \begin{equation}
   I(r_{\rm{e}}) = \frac{f_{\rm{R}}}{K_{\rm{\beta}} r^{2}_{\rm{e}} \left(1-\epsilon \right)},
   \end{equation}
   where $f_{\rm{R}}$ is the total flux of the galaxy, $\epsilon$ is the ellipticity, and $K_{\rm{\beta}}$ can be derived from
   \begin{equation}
   K_{\rm{\beta}}=dex \left(0.030 \log^2 \beta - 0.441 \log \beta+1.079 \right),
   \end{equation}
   where $dex$ means $dex(x)=10^x$. Equations (1) to (4) indicate that if we obtained the parameters of $\beta$,
   $\epsilon$, $r_{\rm{e}}$, and $f_{\rm{R}}$, we could confirm the surface brightness [$I(r)$] distribution of the host
   galaxy. Combining with the position angle $\theta$, we can simulate the host of Mrk 501 in the observed images.
   However, the lower resolution and the relatively poor S/N restrict us to measure the accurate values of the above
   parameters. Fortunately, Nilsson et al. (1999) has obtained all the above parameters from the high-resolution images
   in the $R$ band. The free $\beta$ + core model was adopted in our simulations [based on the properties of BL Lac
   objects and the de Vaucouleurs model (e.g., Makino et al. 1990)]. We simulated the host component of Mrk 501, and
   convolved the simulation results into 28 different FWHMs with the point spread function (PSF) of Gaussian profile. The
   FWHMs of the convolved images are from 0.5 to 5.9 arcsec with a bin size of 0.2 arcsec. We performed the photometry
   using 111 fixed apertures from 1.0 to 12.0 arcsec with a bin size of 0.1 arcsec. Table 6 shows the flux simulations for
   the host galaxy under different FWHMs and apertures. Figure 3 shows the relationship among the brightness, FWHMs,
   and apertures. Our simulation results are very different from those in Nilsson et al. (2007). The host subtraction based
   on the subtraction data in Nilsson et al. (2007) led to a significant seeing-brightness correlation for Mrk 501 (see an
   example presented in Figure 2 in Feng et al. 2017). Thus, a reasonable host subtraction is needed for the optical
   photometry for Mrk 501.
\begin{table}
\begin{center}
\begin{minipage}[]{100mm}
\caption[]{The simulation data for host galaxy of Mrk 501}
\label{Table6}
 \end{minipage}
 \setlength{\tabcolsep}{2.5pt}
\small
\begin{tabular}{cccccccccccccccc}
\hline\noalign{\smallskip}
  Apert  &   &   &   &  &   &   & FWHM &  &   &  &  &   &   &   &   \\ \cline{2-16}

   & 0.5  & 0.7  & 0.9   & 1.1 & 1.3  & 1.5   & 1.7  & 1.9  & 2.1 & 2.3 & 2.5 & 2.7 & 2.9  & ... &  5.9 \\

\hline\noalign{\smallskip}
1.0   & 2.615  & 2.138  & 1.732  & 1.425  & 1.190  & 1.012  & 0.872  & 0.762  & 0.672  & 0.599  & 0.537  & 0.486  & 0.441  & ... & 0.164 \\
 1.1   & 2.928  & 2.456  & 2.023  & 1.681  & 1.413  & 1.207  & 1.043  & 0.913  & 0.806  & 0.719  & 0.646  & 0.585  & 0.532  & ... & 0.199 \\
 1.2   & 3.231  & 2.775  & 2.322  & 1.949  & 1.649  & 1.415  & 1.226  & 1.076  & 0.951  & 0.850  & 0.764  & 0.692  & 0.630  & ... & 0.236 \\
 1.3   & 3.512  & 3.082  & 2.620  & 2.222  & 1.893  & 1.632  & 1.419  & 1.249  & 1.106  & 0.990  & 0.891  & 0.808  & 0.736  & ... & 0.277 \\
 1.4   & 3.782  & 3.383  & 2.920  & 2.503  & 2.147  & 1.860  & 1.623  & 1.431  & 1.271  & 1.139  & 1.026  & 0.931  & 0.849  & ... & 0.321 \\
 1.5   & 4.034  & 3.668  & 3.213  & 2.783  & 2.405  & 2.095  & 1.834  & 1.622  & 1.443  & 1.296  & 1.168  & 1.062  & 0.969  & ... & 0.368 \\
 ...   & ...    & ...    & ...    & ...    & ...    & ...    & ...    & ...    & ...    & ...    & ...    & ...    & ...    & ... & ...   \\
 12.0  &15.554  &15.542  &15.527  &15.509  &15.486  &15.460  &15.429  &15.394  &15.354  &15.311  &15.259  &15.206  &15.143  & ... &12.889 \\
\noalign{\smallskip}\hline
\end{tabular}
\end{center}
 \tablecomments{0.86\textwidth}{This table is available in its entirety in a machine-readable form in the online journal. A portion is shown here for
guidance regarding its form and content. Apert: aperture radius in units of arcsec, presented in column 1. FWHM is in units of arcsec, presented in columns 2--16. The fluxes are in unit of mJy.}
\end{table}
   \begin{figure}
  \begin{center}
   \includegraphics[width=0.45\textwidth]{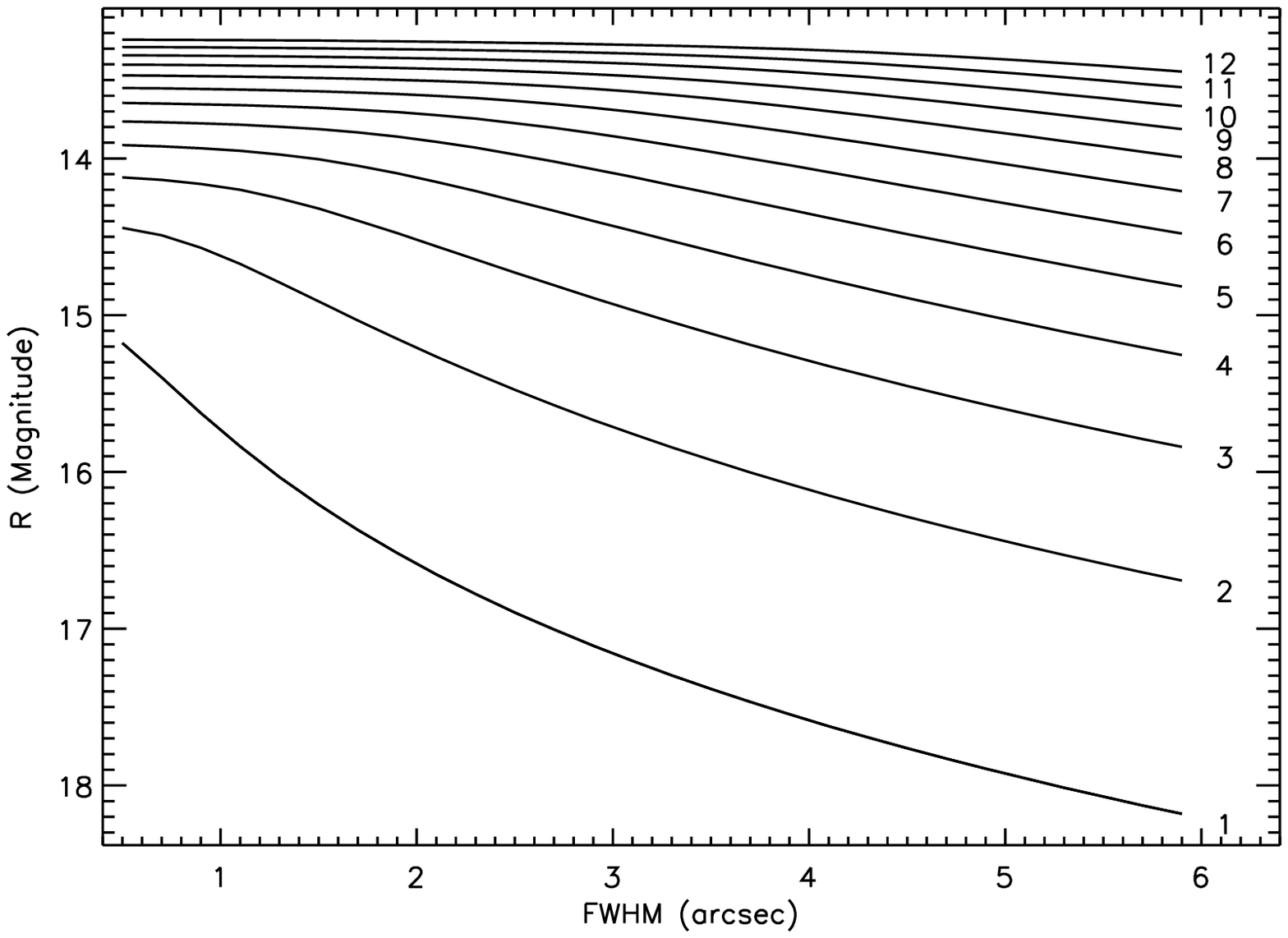}
   \includegraphics[width=0.45\textwidth]{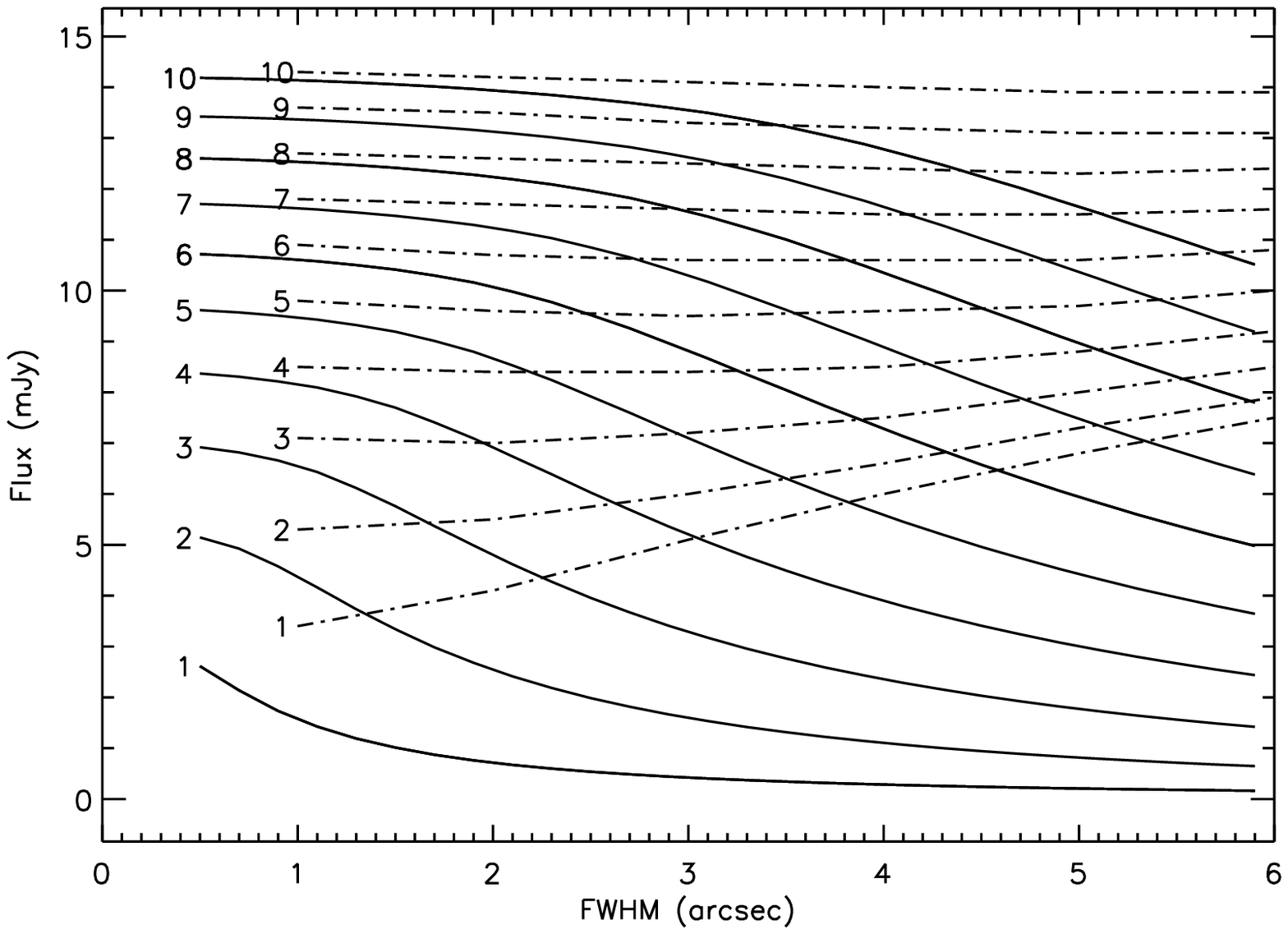}
  \end{center}
  \caption{The relationships between FWHM and brightness for different apertures in simulation to the host galaxy of Mrk 501
     (the solid lines). The dotted-dashed lines represent the results of Nilsson et al. (2007). The numbers following the lines in plots
     are the photometric aperture radii.}
  \label{fig3}
  \end{figure}

   We used two methods to compare the simulation results with our observations. First, we checked the observed images to
   determine photometric regions where the S/N ratios are high enough (i.e., $>$ 5). This is normally achieved with an aperture
   radius of 5 arcsec. We measured the brightness of the images within annular apertures with radii of 3.5--4.5 and 4.5--5.0 arcsec.
   Figure 4 shows the comparisons between the simulated and observed results in the same annular apertures. The simulations
   and observations are (marginally) consistent with each other in the case of 3.5--4.5 arcsec except for 2010 May 17 (see Figure 4).
   In general, the observed results are less than the simulation results in the case of 4.5--5.0 arcsec. This may rise from low S/N
   ratios at those annular apertures. The host galaxy of Mrk 501 is a low surface brightness galaxy, and this will result in lower
   S/N ratios at larger annular apertures. The deviations of simulations from observations in the case of 3.5--4.5 arcsec are less
   than those in the case of 4.5--5.0 arcsec. Combing four panels in Figure 4 into one panel (see Figure 5), we find that simulations
   are marginally consistent with observations in the case of 3.5--4.5 arcsec, and the deviations of observations from simulations
   in the case of 3.5--4.5 arcsec are less than those in the case of 4.5--5.0 arcsec. Observations need an exposure time to obtain a
   certain S/N ratio. A low S/N ratio may result in a lower flux measurement compared to the flux simulation based on a high S/N
   ratio image presented in Nilsson et al. (1999). Another method is based on the fact that the brightness difference between
   simulation and observation is the contribution of AGN, i.e., the observed flux is a combination of AGN and its host galaxy flux,
   while the simulation result only contains the host component. For a relatively large photometric aperture (nearly including all
   the AGN flux, e.g., an aperture radius of 4.0 arcsec including 99\% of the AGN flux), the differences between simulations and
   observations should be a constant for the different seeing conditions. The observed results are well consistent with the vertically
   shifted simulation results for the aperture radii from 3.0 to 6.0 arcsec in the flux versus FWHM diagram (see Figure 2). There are
   very similar trends between the vertically shifted simulations and the observational results for the other aperture radii in Figure 2.
   These slight differences between simulations and observations may be from the fact that the corresponding aperture radii are
   either too small or too large ($<$3.0 or $>$6.0 arcsec).
  \begin{figure}
 \begin{center}
  \includegraphics[angle=0,scale=0.4]{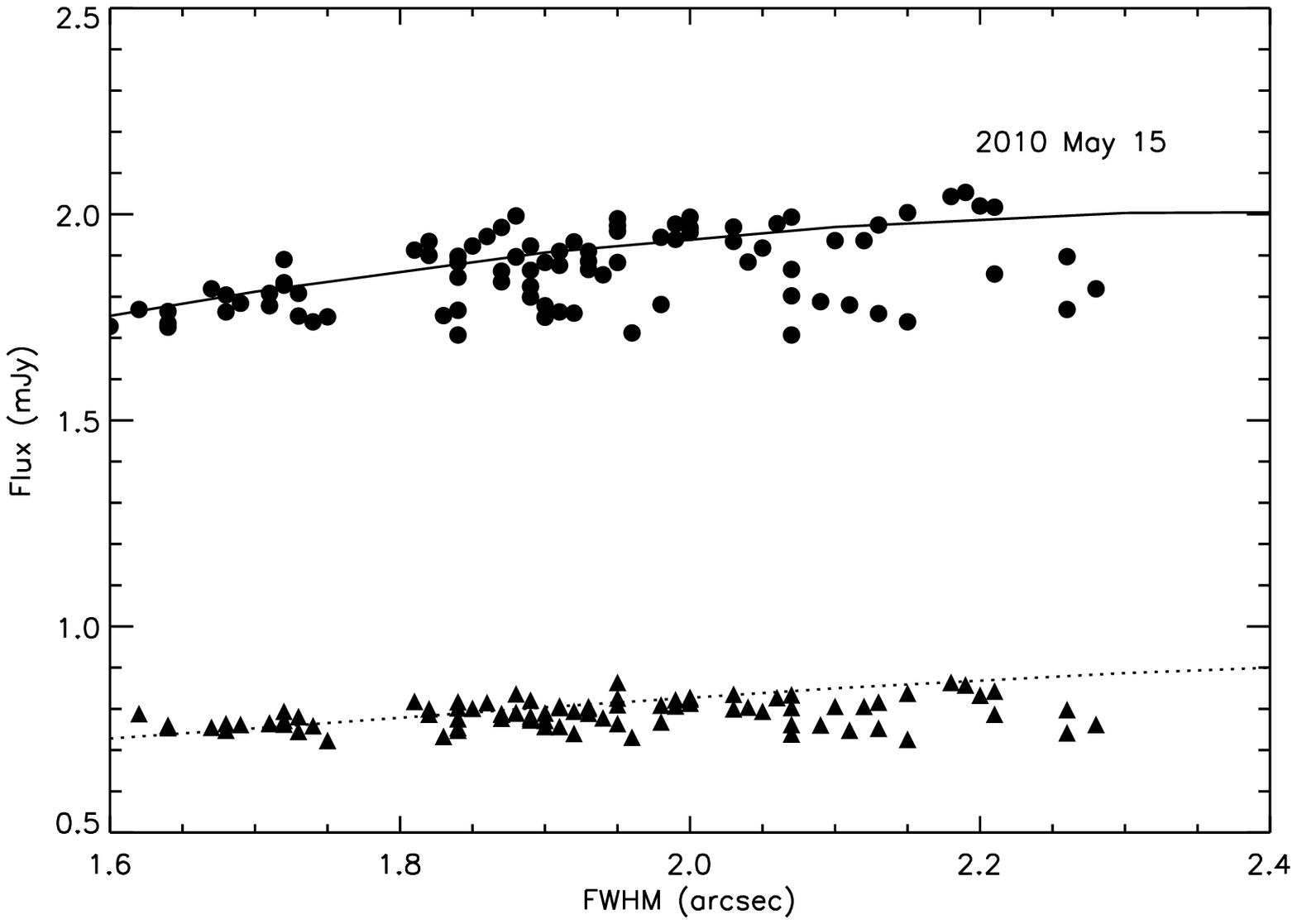}
  \includegraphics[angle=0,scale=0.4]{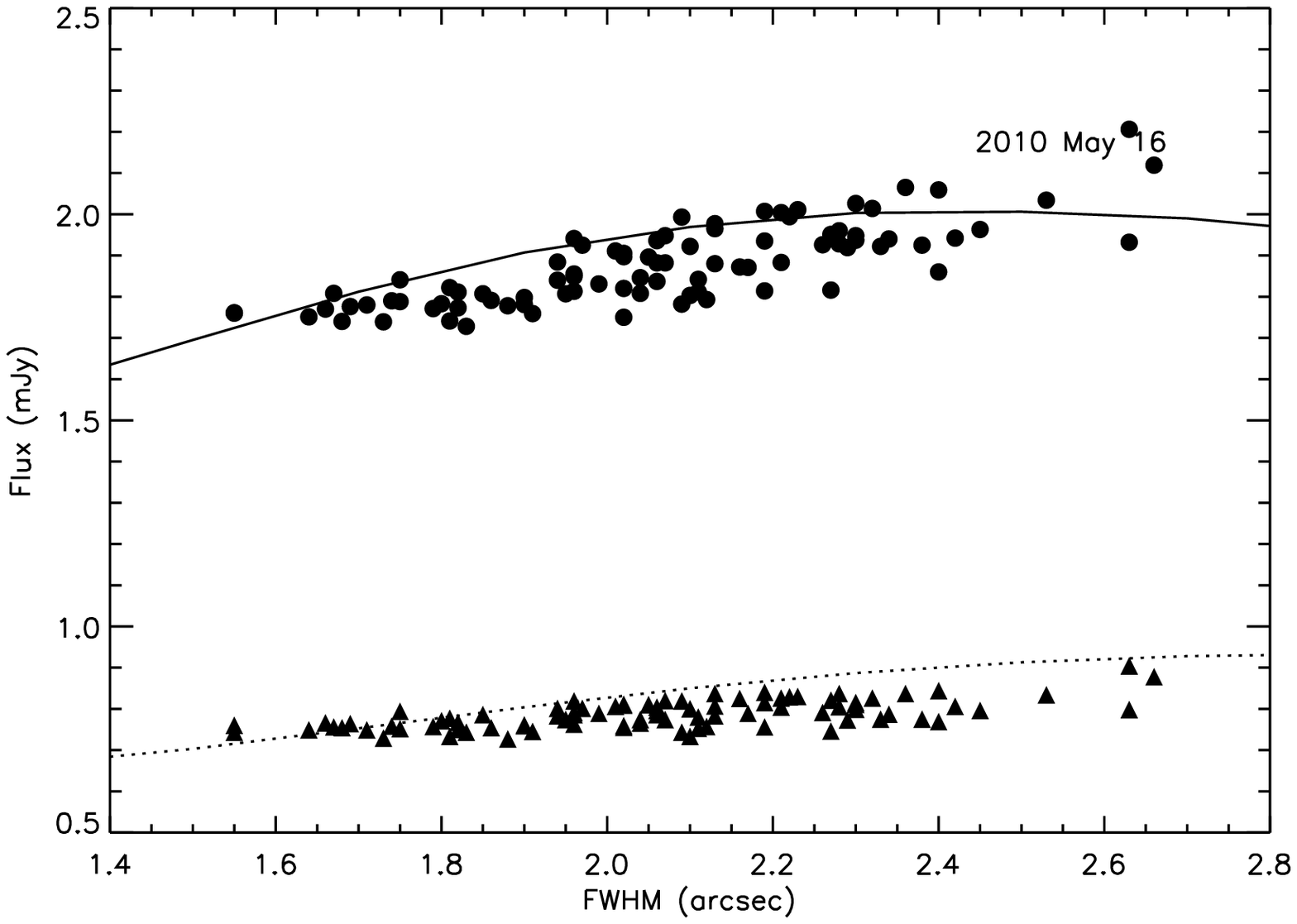}
  \includegraphics[angle=0,scale=0.4]{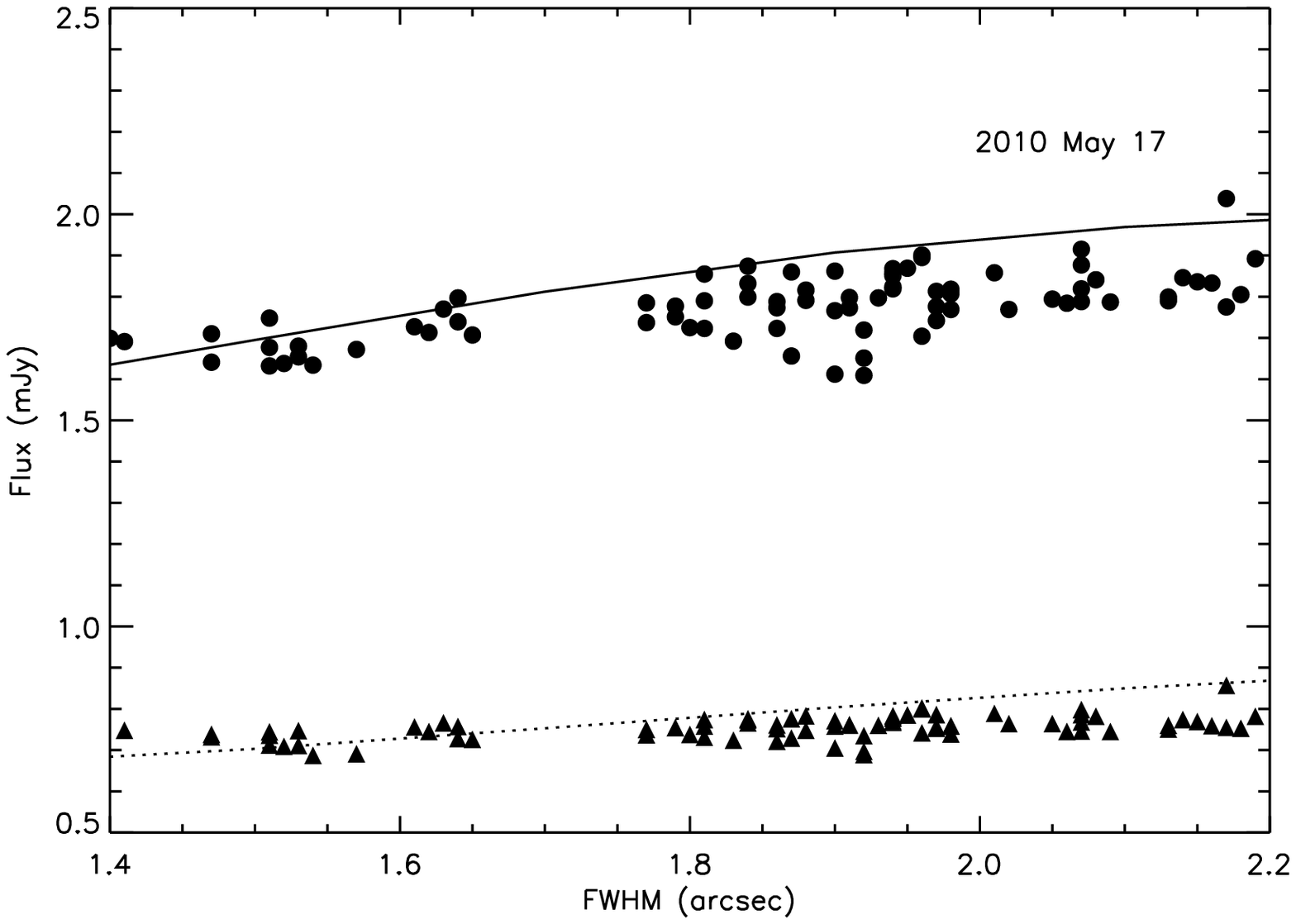}
  \includegraphics[angle=0,scale=0.4]{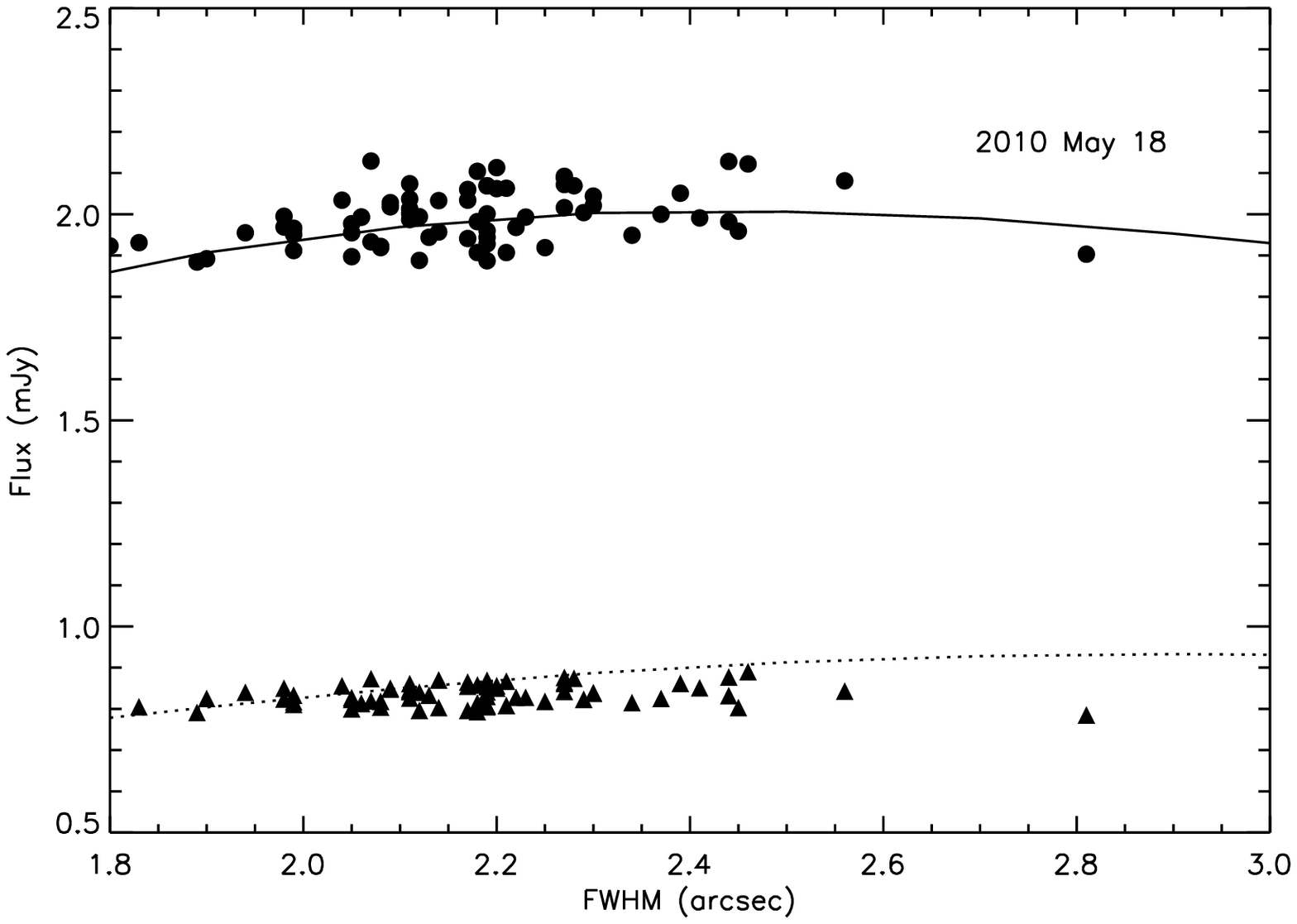}
 \end{center}
 \caption{Fluxes in annular apertures for different seeing (FWHM). In each panel, the solid line denotes the simulation results from
    3.5 to 4.5 arcsec, and the circles denote the observed results in the same annular apertures. In each panel, the dotted line denotes
    the simulation results  from 4.5 to 5.0 arcsec, and the triangles denote the observed results in the same annular apertures.}
  \label{fig4}
\end{figure}
   \begin{figure}
   \begin{center}
        \includegraphics[angle=0,scale=0.4]{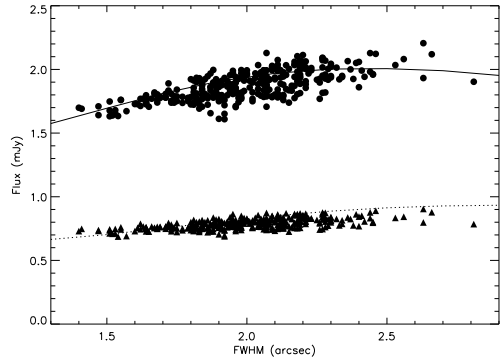}
     \end{center}
     \caption{These observational fluxes on 2010 May 15 -- 18. The symbols are same as in Figure 4.}
  \label{fig5}
\end{figure}

   We calculated the average difference between the simulations and the observational results in Figure 2. The shifted simulation
   results are well consistent with the observational data, and the average difference can be regarded as the AGN flux. The mean
   flux of AGN is $\sim$ 8.0 mJy, which corresponds to $R\sim13^{m}_{\cdot}96$ [$F = 3.08 \times 10^{-0.4 \times R+3}$ $\rm{Jy}$
   (Nilsson et al. 2007)]. Thus, AGN's contribution to the total flux of the source is $\sim$ 13.3\%. Compared with the brightness
   obtained in Nilsson et al. (1999), $R=14^{m}_{\cdot}45$, AGN Mrk 501 brightened by $\sim$ 57\% in our observations. According
   to our simulations, we subtracted the host contribution, and investigated whether there are still significant seeing-brightness
   correlations for AGN Mrk 501. The host-subtracted fluxes versus FWHMs are presented in Figure 6. There is no correlation on
   2010 May 18. Though the host-subtraction based on our simulations can (obviously) weaken the significant correlations found
   in Feng et al. (2017), there are still correlations for 2010 May 15 and 17, and an obvious correlation on 2010 May 16. Figure 7
   shows the host-subtracted flux versus seeing distribution in the case of 5.0 arcsec aperture. The host-subtracted flux versus seeing
   distribution shows that the larger photometric aperture radii can further weaken the host-subtracted brightness-seeing correlation.
   Thus, our simulations can basically give a reasonable host-subtraction. The obvious correlation on 2010 May 16 might be from
   the smaller photometric aperture relative to the average seeing.  The host-subtracted flux light curves show that the darkening
   variations found in Feng et al. (2017) still exist in the light curve on 2010 May 15 even though the host contribution has been
   subtracted (see Figure 8). There is a flare with a duration $\sim$ 1 hours on 2010 May 18 around MJD 5334.75 (see Figure 8), which
   was not found in Feng et al. (2017). This confirms that the fake large amplitude fast variability due to the seeing effect can mask
   the intrinsic micro variability in Mrk 501. This kind of fake rapid and strong variability due to seeing effect will mask the intrinsic
   micro variability in Mrk 501, and will lead to difficulty in detecting the intrinsic micro variability in similar sources with brighter
   host galaxies, e.g., Mrk 421.
   \begin{figure}
   \begin{center}
     \includegraphics[angle=0,scale=0.4]{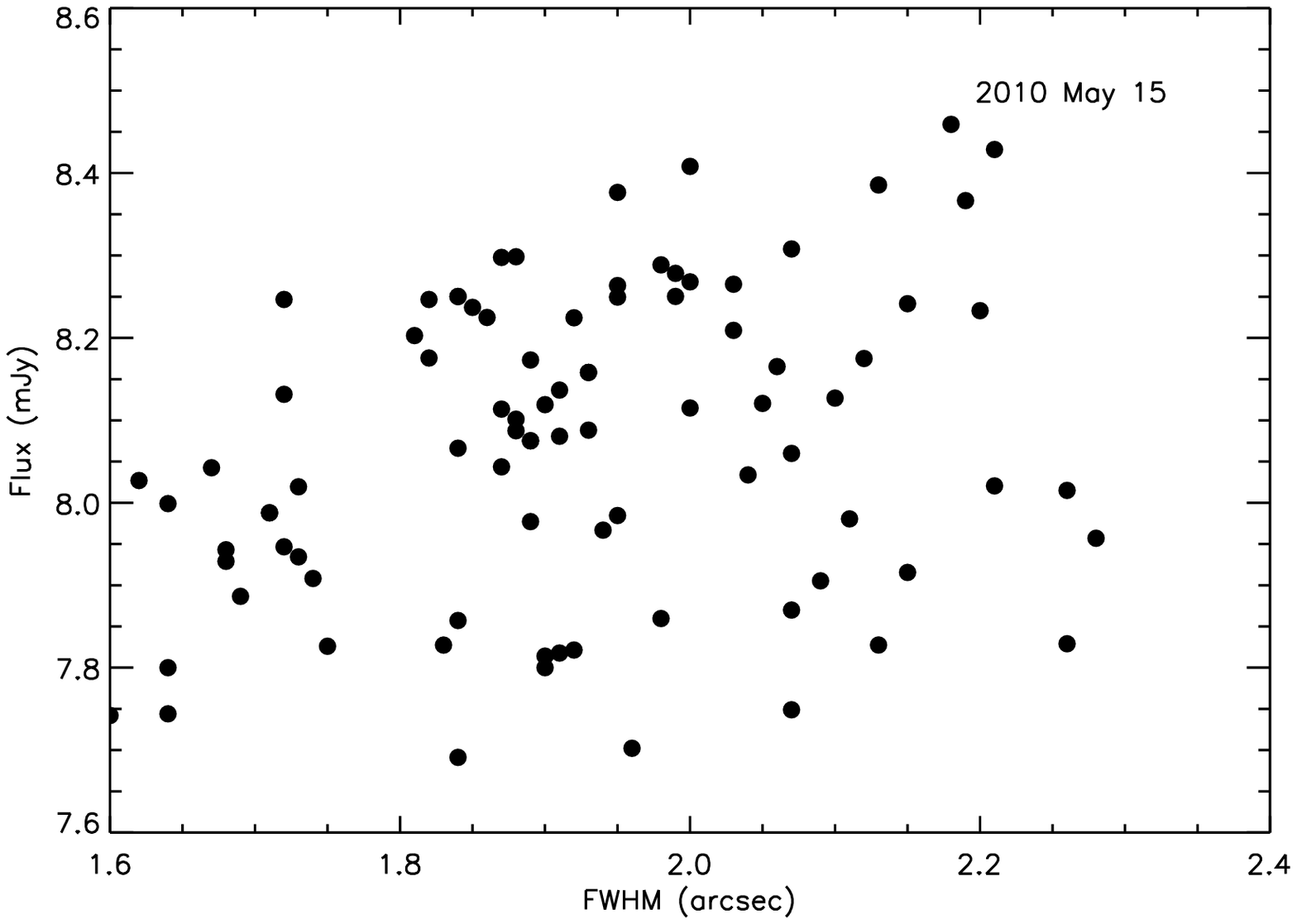}
     \includegraphics[angle=0,scale=0.4]{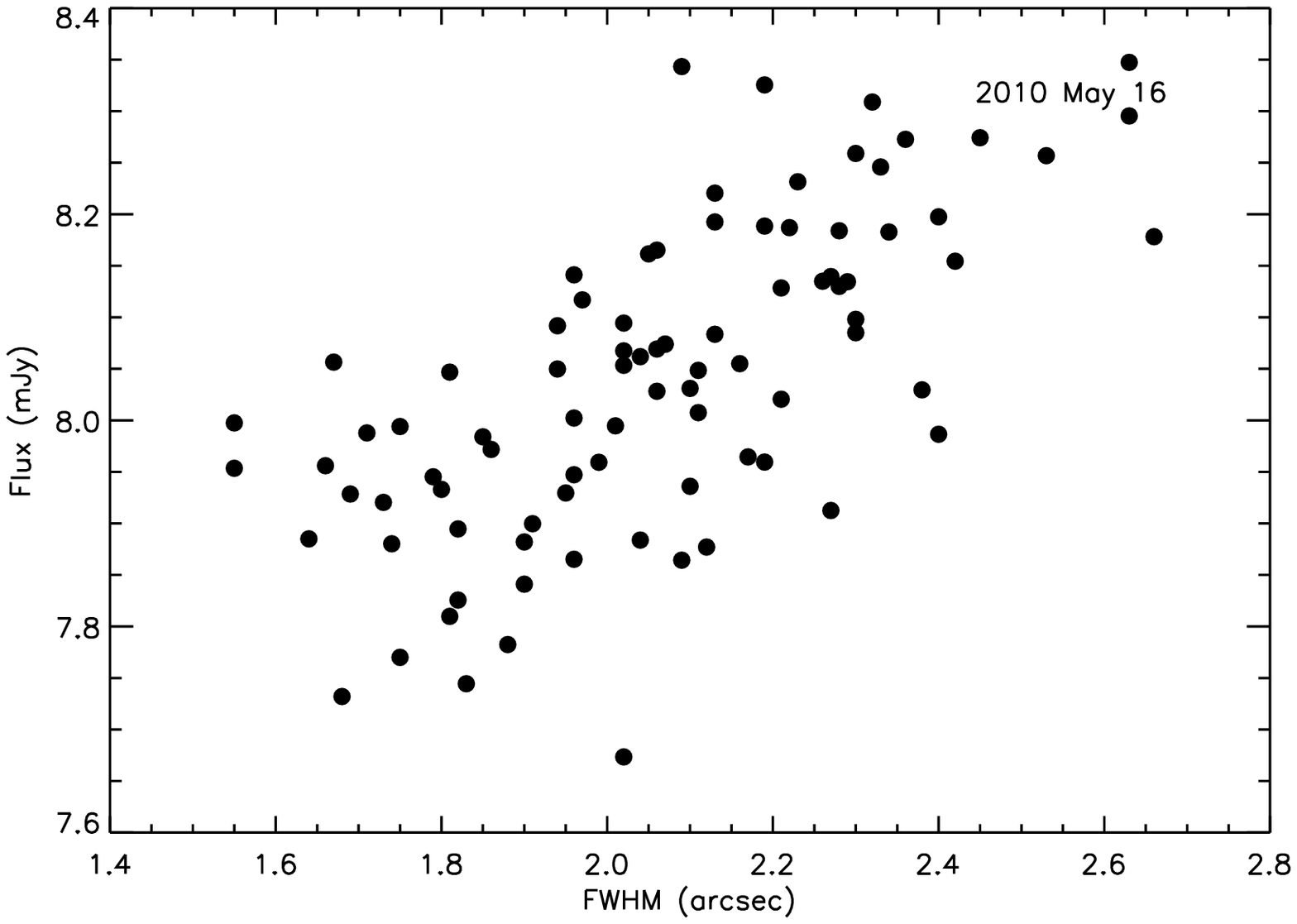}
     \includegraphics[angle=0,scale=0.4]{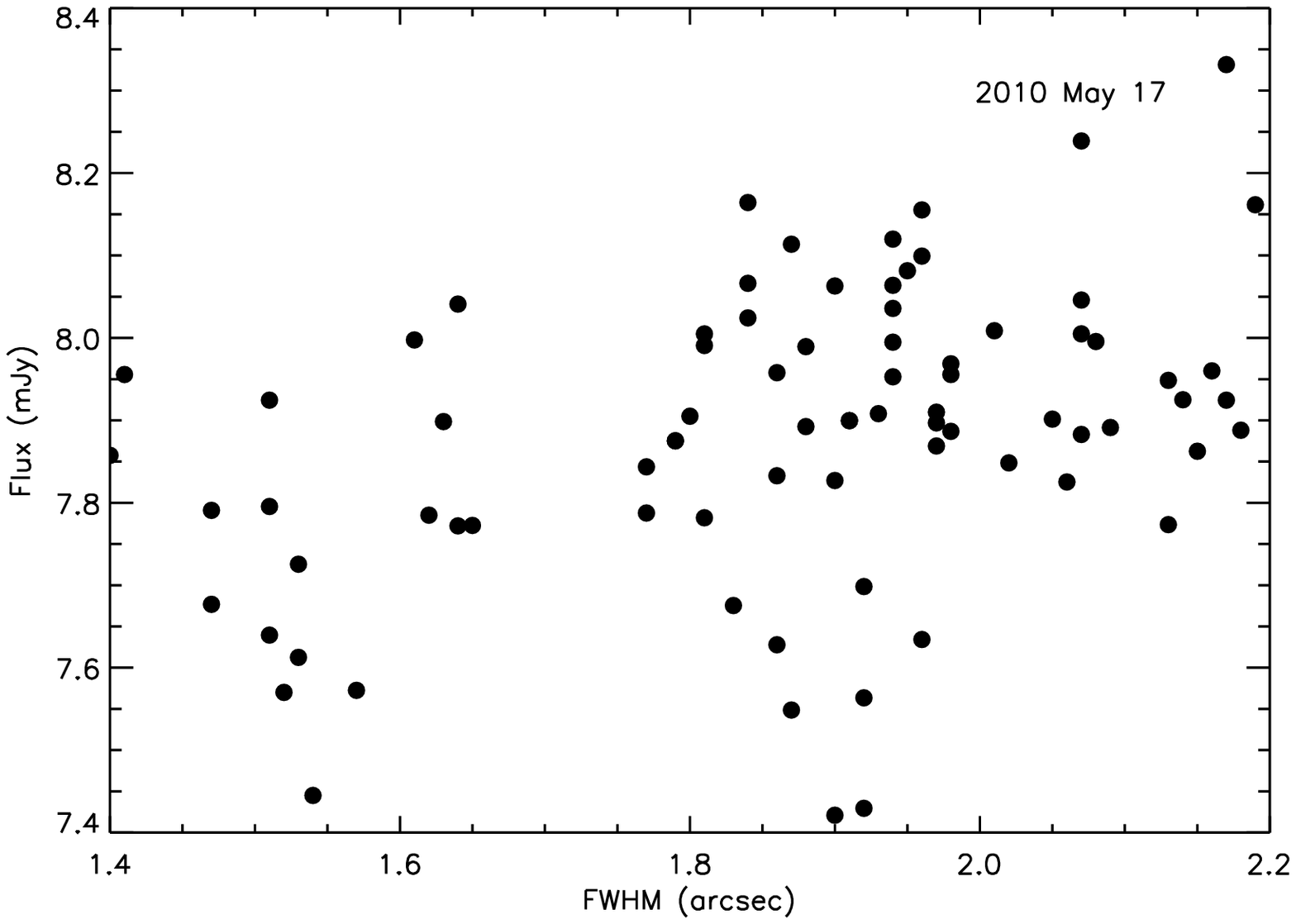}
     \includegraphics[angle=0,scale=0.4]{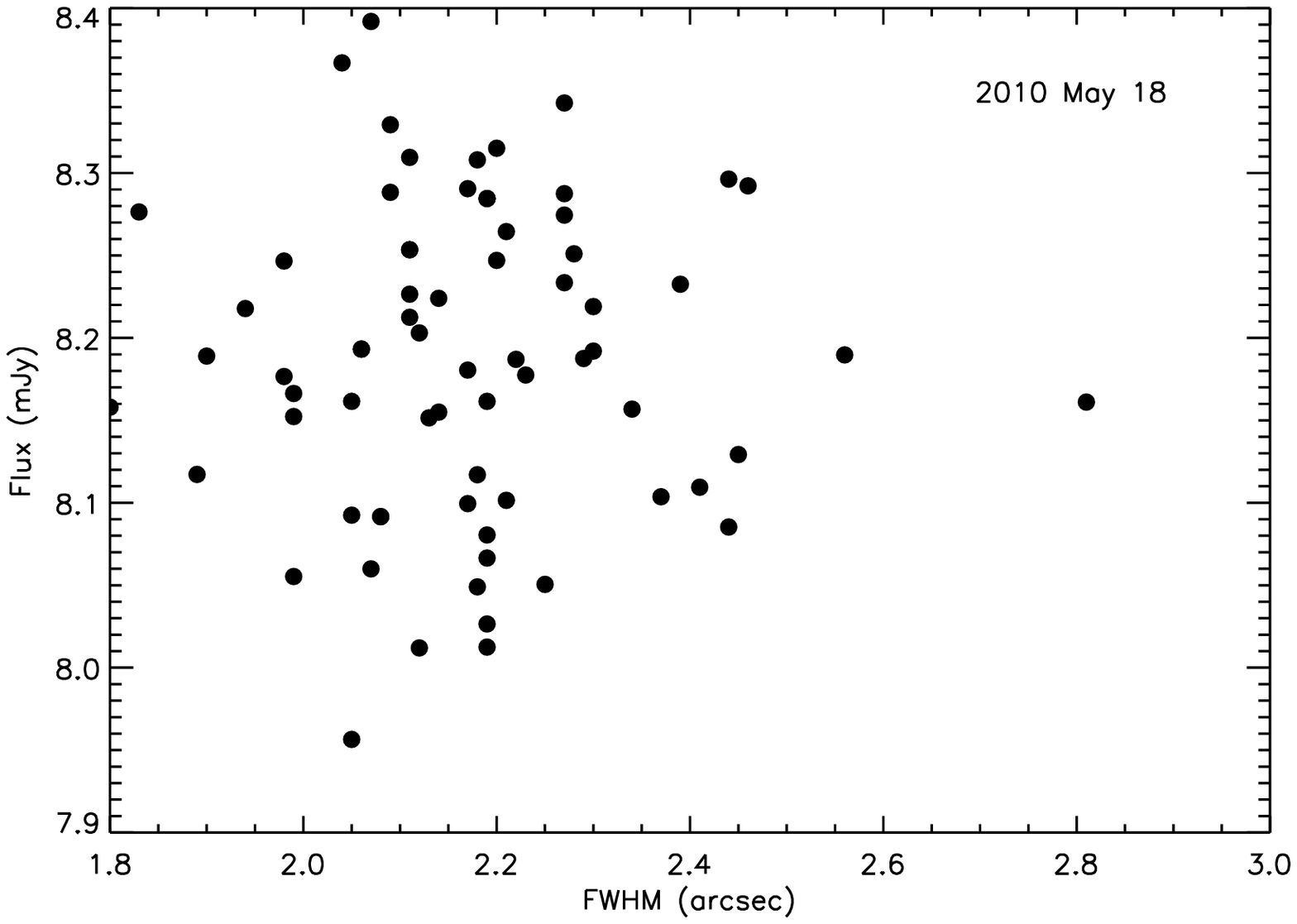}
    \end{center}
  \caption{AGN host-subtracted flux versus seeing (FWHM). The observed fluxes are measured for the photometric
     aperture radius of 4.0 arcsec.}
  \label{fig6}
\end{figure}
   \begin{figure}
   \begin{center}
        \includegraphics[angle=0,scale=0.4]{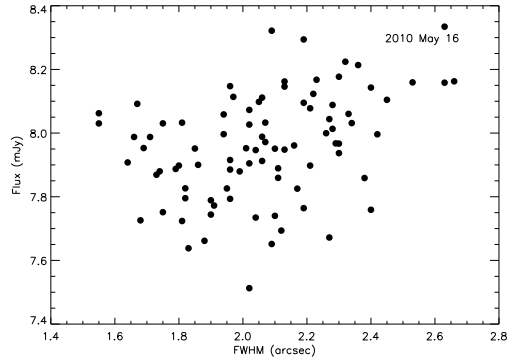}
    \end{center}
     \caption{AGN host-subtracted flux versus seeing (FWHM) on 2010 May 16 for an aperture radius of 5.0 arcsec.}
  \label{fig7}
\end{figure}

   \begin{figure}
     \begin{center}
        \includegraphics[angle=0,scale=0.4]{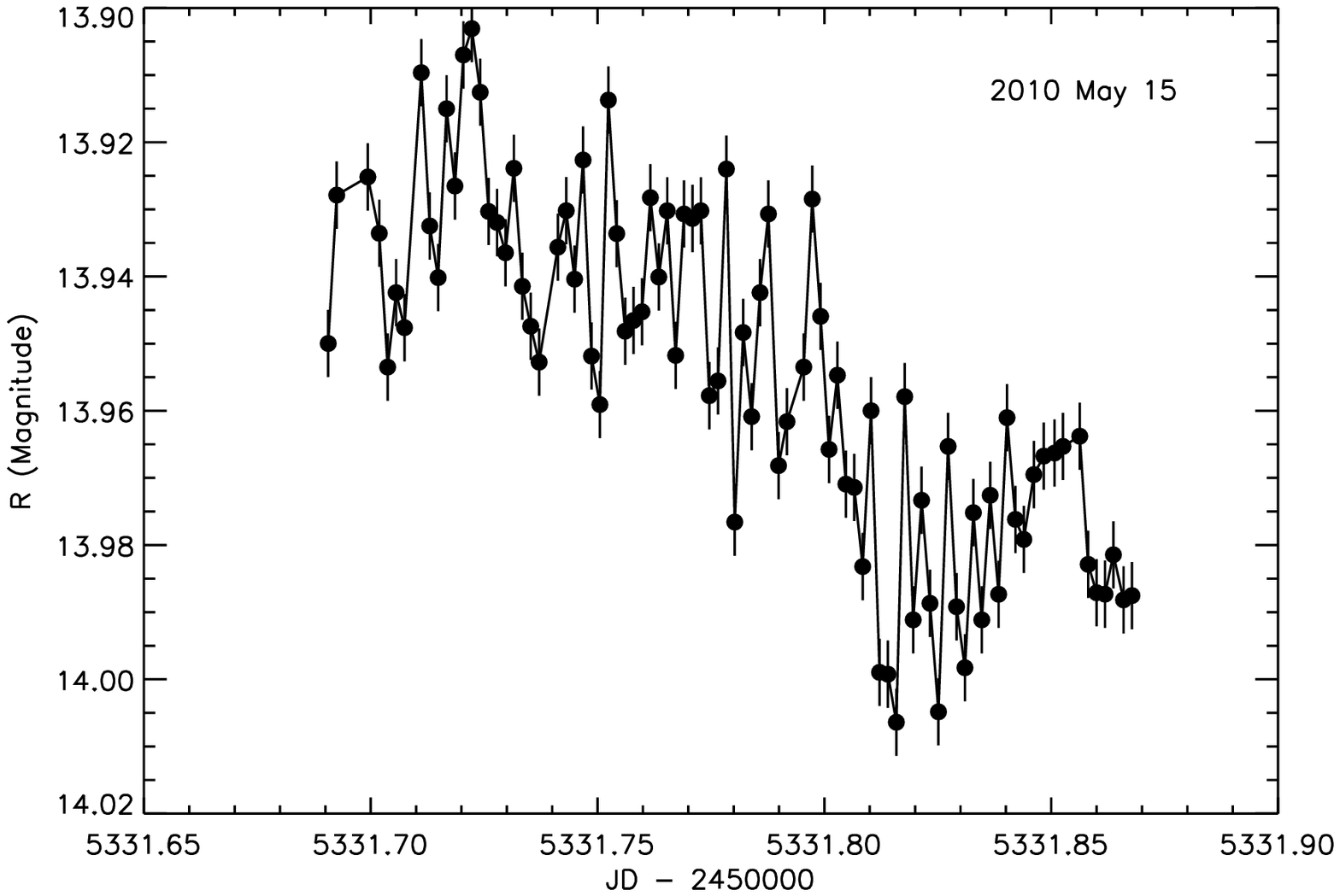}
        \includegraphics[angle=0,scale=0.4]{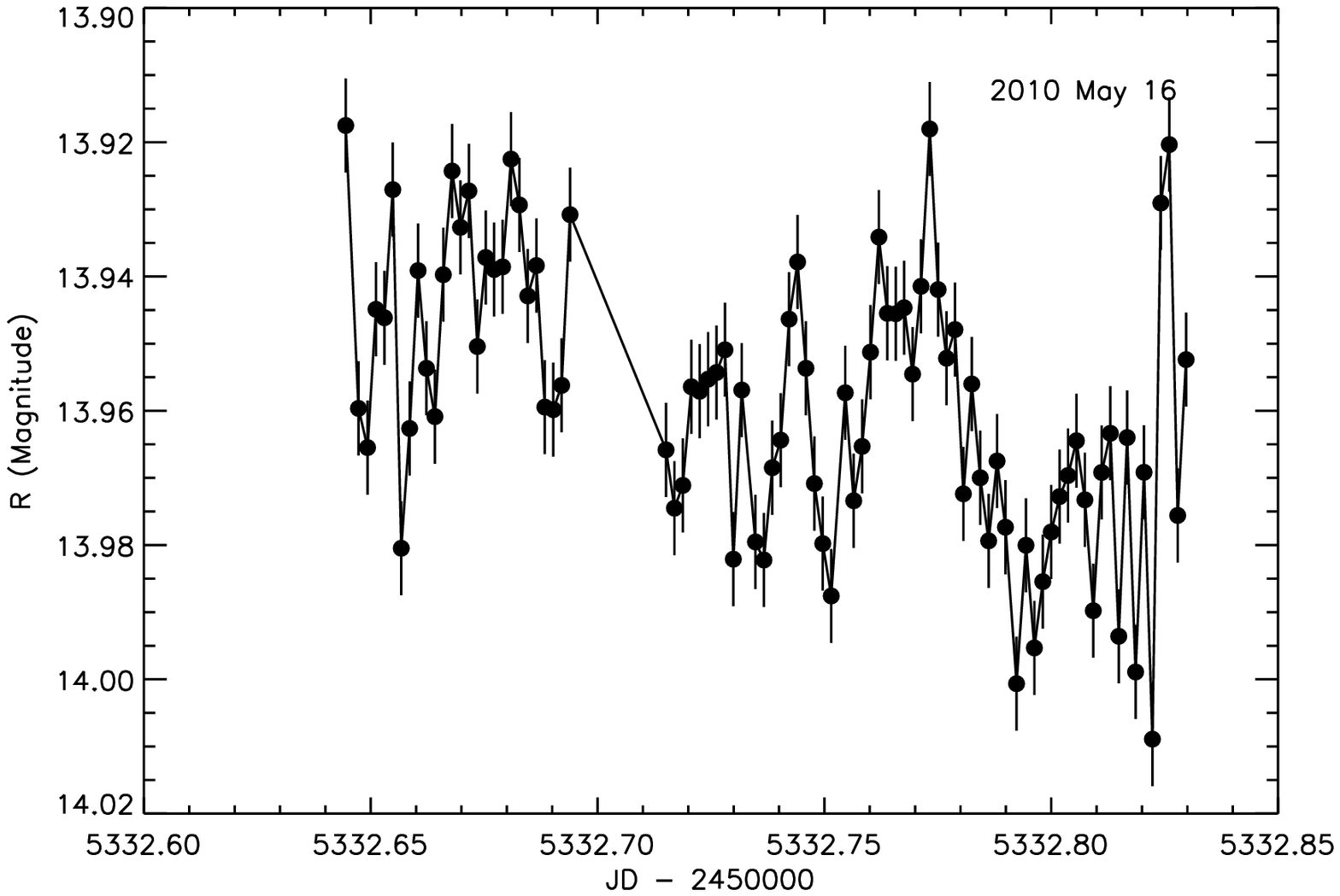}
        \includegraphics[angle=0,scale=0.4]{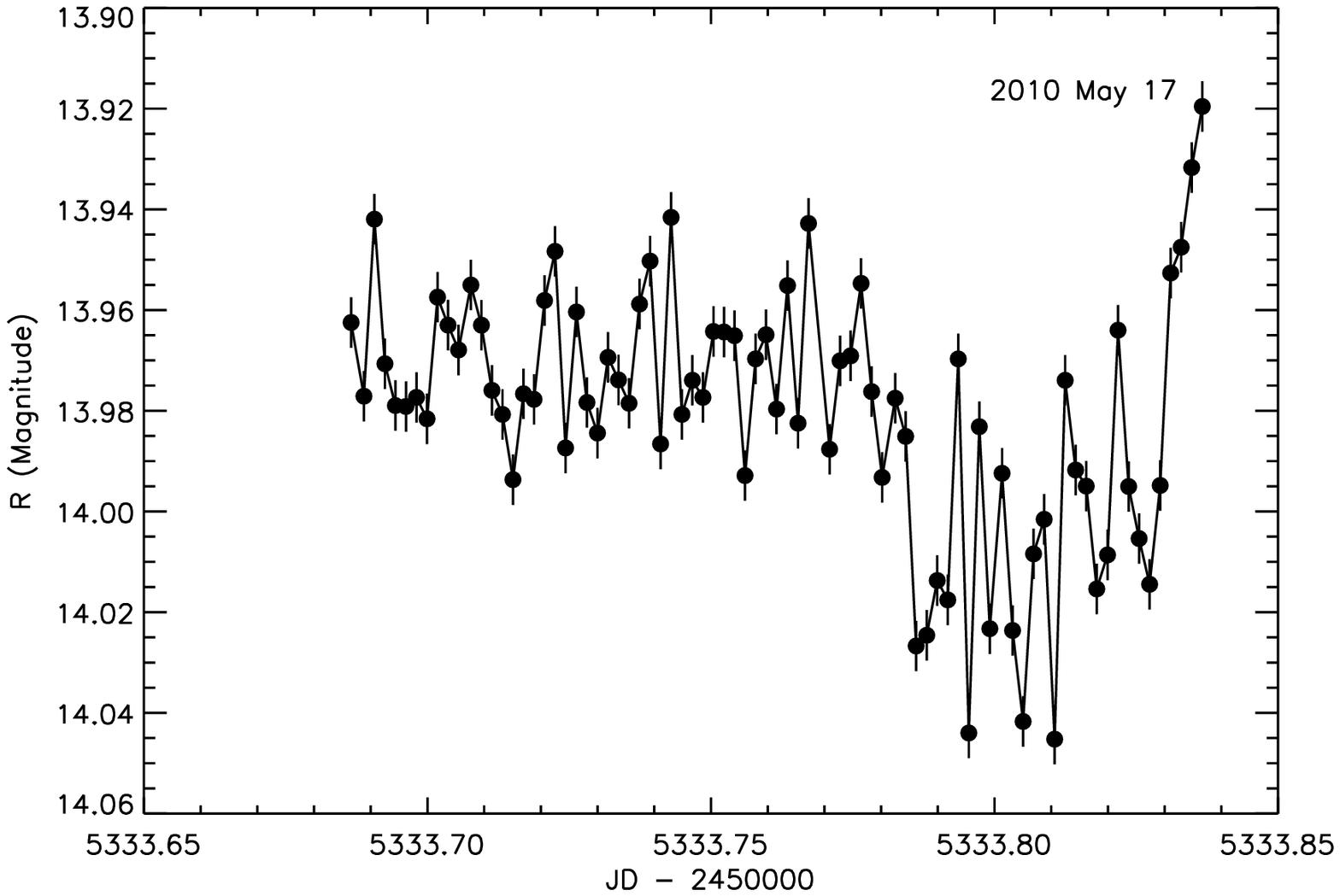}
        \includegraphics[angle=0,scale=0.4]{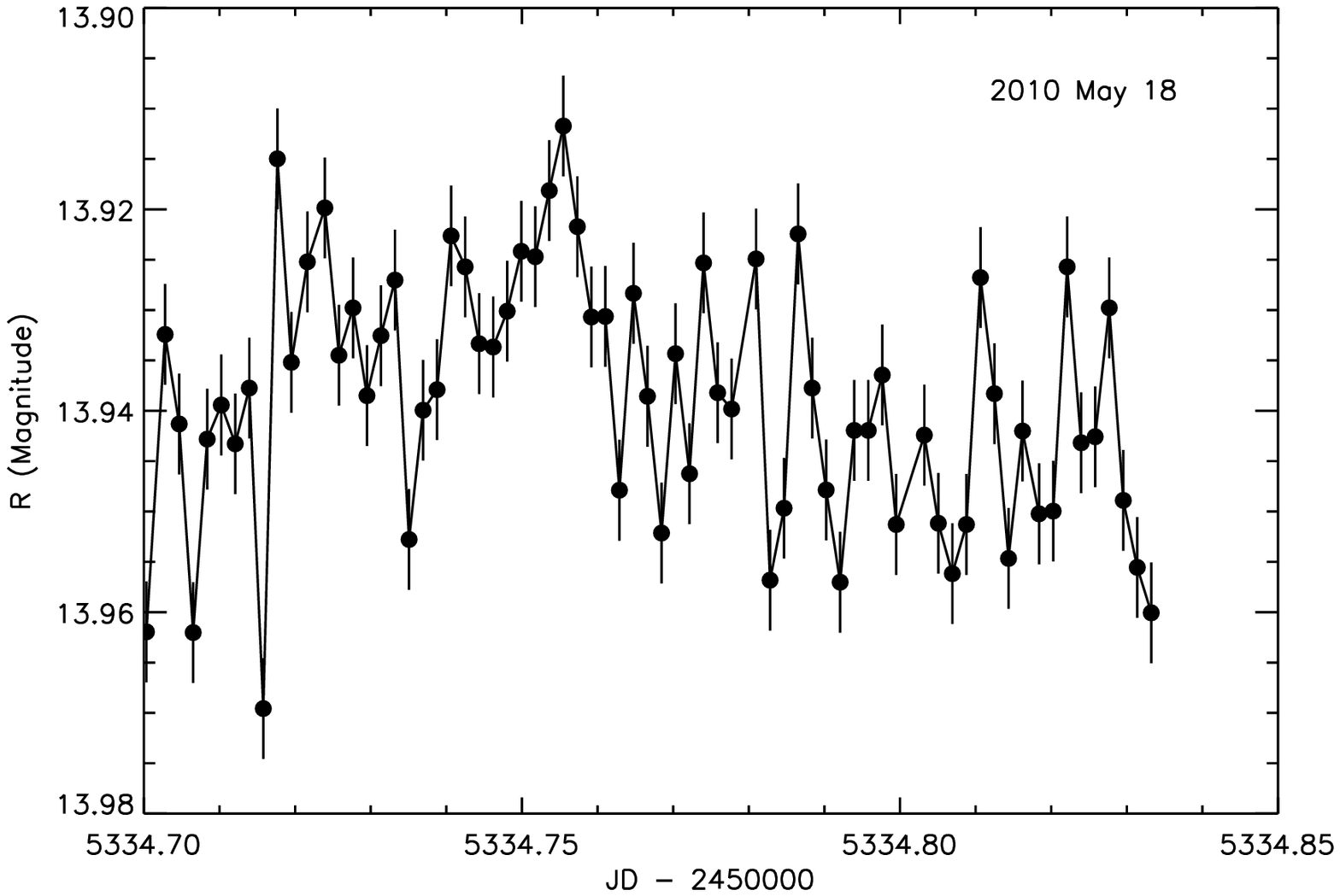}
       \end{center}
     \caption{AGN host-subtracted light curves for a photometric aperture radius of 4.0 arcsec. }
     \label{fig8}
   \end{figure}

\section{Conclusions}
  Based on the intensive observations run with the 1.02 m optical telescope at Yunnan Observatories from 2010 May 15 to 18,
  and a two-dimensional model of elliptical galaxy, we simulated the $R$-band contribution of the host galaxy of TeV $\gamma$-ray
  BL Lac object Mrk 501. The simulated brightness in the different aperture radii and seeing conditions shows correlations between
  the seeing and brightness for the host galaxy, and these correlations are well confirmed by the observational data. The differences
  between the simulation fluxes and the observational data are due to AGN Mrk 501 contribution, and the host-subtracted brightness
  of Mrk 501 can obviously weaken these significant correlations found in Feng et al. (2007). There is no correlation between the seeing
  and the host-subtracted brightness on 2010 May 18. However, there are correlations on 2010 May 15 and 17, and an obvious correlation
  on 2010 May 16. The larger photometric aperture radii with respect to the seeing average can further weaken the correlation on 2010
  May 16 (see Figures 6 and 7). These correlations led to illusive large amplitude variations on short timescales, which can mask the
  intrinsic micro variability and then lead to the difficulty in detecting the intrinsic micro variability. The host-subtracted brightness
  light curves confirm the darkening variations on 2010 May 15 found in Feng et al. (2017), and discover a flare with a duration $\sim$
  1 hours on 2010 May 18. Both of the aperture size and the seeing condition influence the photometric results, but the aperture size
  can generate a more serious influence. The pure nuclear flux is $\sim$ 8.0 $\rm{mJy}$. Compared with the result observed in 1996 July
  (Nilsson et al. 1999), the AGN Mrk 501 brightened by a factor of $\sim$ 57\%. Simulation data of the host galaxy of Mrk 501 are given
  for the different aperture radii and seeing conditions (on-line Table 6).

\section{Discussion}
  The correlation between the seeing FWHM and the brightness within a certain aperture is obvious for the intensive observations on
  2010 May 15 to 18. At the same time, the flux of the target is higher as the aperture radius is larger. The larger aperture radius will
  cover a more area of an extended source, and then will contain more light in the aperture. Thus, the total brightness will be monotonously
  increasing with the aperture radius. This indicates that a fixed aperture is better than a dynamic aperture in performing photometry
  for Mrk 501. This point was tested in Feng et al. (2017), where a fixed aperture was used to perform photometry. Brightness monotonously
  decreases with the increasing FWHM of seeing in the fixed aperture. This can be explained that the larger PSF due to the worse seeing
  will scatter out more light from a fixed aperture. Another feature is that the PSF effect is more significant for the smaller aperture (less
  than 3.0 arcsec). This is due to the fact that the scattered light of AGN changes for the different PSFs. Therefore, the photometry of Mrk 501
  should use a large fixed aperture, which can contain almost all the light of the AGN. In addition, it is necessary to correct the influence
  of seeing.

   Figure 3 shows the simulation results of the host galaxy of Mrk 501,  and the two panels in Figure 3 present the similar relationships in
   Figures 1 and 2. The brightness curve shapes of simulation results are very similar to those of observations for the same aperture and
   the same range of FWHM. However, the results in Figure 3 are somewhat different from the results in Figures 1 and 2, especially in the
   small apertures, and this difference is mainly due to the AGN component. We tested the reliability of the simulations via two methods
   (see Section 3), and both tests indicate that the simulations are robust (see Figure 2). The results in Figure 3 can be used to correct the
   host contamination of Mrk 501, and the corresponding values are given in Table 6. Nilsson et al. (2007) had given a similar table (Table B.1). Comparing our simulation results to theirs (see Figure 3), we found some differences. Though these two results indicate that the host
   fluxes depend strongly on the photometric apertures, the values from the same aperture and PSF are inconsistent. Especially within
   small aperture radius ($\leq$ 3.0 arcsec), the difference is significant. For a fixed aperture, the relationships between brightness and
   FWHM are significantly different for these two results. The brightness of the host galaxy decreases as the FWHM increases (see Figure 3).
   These trends are opposite to the results in Nilsson et al. (2007). The influence of the seeing on the variability amplitude is significant
   in our results. After we subtracted the contamination of the host galaxy using the results of Nilsson et al. (2007), the relationships are
   still significant for the brightness and seeing FWHM. Otherwise, if the brightness of the host galaxy monotonously increases with the
   FWHM, the outer part of the host galaxy would be brighter than the central part. This is inconsistent with the universal of the surface
   brightness distribution of elliptical galaxy.

   The simulations and observations indicate that the AGN contribution of Mrk 501 is $\sim$ 13.3\%. This means that even the variable
   of AGN is up to 10\%, we can only detect a magnitude change $\sim 0^{m}_{\cdot}01$ for the whole galaxy. This variability amplitude
   approximates to the limit accuracy of photometry for some telescopes. Therefore, it is not easy in detecting this variability in Mrk 501.
   The effects of the photometric aperture and the observational seeing are significant for the photometric results, and most of the previous
   works didn't take into account the effects of the two factors. This might lead to some fake variability in some previous works for Mrk 501,
   and the relevant results should be reconsidered. Our studies suggest that a fixed aperture, which depends on the seeing condition, is better
   than a dynamic aperture, and the host galaxy subtraction is necessary. Our simulations give a reasonable host-subtraction. The strong
   host contamination also impact the color, polarization, and SED of AGN. Thus, it is meaningful to subtract the host component before
   investigating the property of Mrk 501.

\begin{acknowledgements}
 We are grateful to the anonymous referee for constructive comments leading to significant improvement of this paper, and the editor for
 helpful suggestions. We thank the financial supports of the Key Research Program of the CAS (Grant No. KJZD-EW-M06), the National Natural Science Foundation of China (NSFC; Grant No. 11433004), and the Ministry of Science and Technology of China (2016YFA0400700). We also thank the financial supports of the NSFC (Grants No. 11273052 and U1431228), and the Youth Innovation Promotion Association, CAS.

\end{acknowledgements}


\begin{thebibliography}{99}\label{thebibliography}
%% you can type \apj for ApJ, \aap for A&A, \apss for Ap&SS, etc. Please consult
%% the macro chjaa.cls. You can also find them in aasguide.tex (AASTeX for ApJ, AJ, PASP)
%% Please follow the format of ChJAA's reference list

\bibitem[Abdo et al. (2010)]{Ab10} Abdo, A. A., Ackermann, M., Agudo, I., et al. 2010, ApJ, 716, 30
\bibitem[Abdo et al. (2011)]{Ab11} Abdo, A. A., Ackermann, M., Ajello, M., et al. 2011, ApJ, 727, 129
\bibitem[Ahnen et al. (2017)]{Ah17} Ahnen, M. L., Ansoldi, S., Antonelli, L. A., et al. 2017, A\&A, 603, A31
\bibitem[Albert et al. (2007)]{Al07} Albert, J., Aliu, E., Anderhub, H., et al. 2007, ApJ, 669, 862

\bibitem[Angel \& Angel (1980)]{An80}  Angel, J. R. P, \& Stockman, H. S. 1980, ARA\&A, 18, 321

\bibitem[Bai et al. (1998)]{Ba98} Bai, J. M., Xie, G. Z., Li, K. H., et al. 1998, A\&AS, 132, 83B
\bibitem[Blandford \& K\"{o}nigl (1979)]{BK79} Blandfor, R., D., \& K\"{o}nigl, A. 1979, ApH, 232, 34

\bibitem[B\"{o}ttcher \& Dermer (2002)]{Bo02}  B\"{o}ttcher, M, \& Dermer, C. D. 2002, ApJ, 564, 86

\bibitem[B\"{o}ttcher (2007)]{Bo07} B\"{o}ttcher, M. 2007, Ap\&SS, 309, 95
\bibitem[Caon et al. (1993)]{Ca93} Caon, N., Capaccioli, M., \& D$^{,}$Onofrio, M. 1993, MNRAS, 265, 1013
\bibitem[Catanese et al. (1997)]{Ca97} Catanese, M., Bradbury, S. M., Breslin, A. C., et al. 1997, ApJ, 487, 143
\bibitem[Catanese \& Sambruna (2000)]{CS00} Catanese, M., \& Sambruna, R. M. 2000, ApJL, 534, L39
\bibitem[Ciprini et al. (2003)]{Ci03} Ciprini, S., Tosti, G., Raiteri, C. M., et al. 2003, A\&A, 400, 487
\bibitem[Ciprini et al. (2007)]{Ci07} Ciprini, S., Takalo, L. O., Tosti, G., et al. 2007, A\&A, 467, 465
\bibitem[Dai et al. (2015)]{Da15} Dai, B. B., Zeng, W., Jiang, Z. J., et al. 2015, ApJS, 218, 18
\bibitem[Dermer \& Schlickeiser (1993)]{DS93} Dermer, C. D., \& Schlickeiser, R. 1993, ApJ, 416, 458
\bibitem[Falomo (1996)]{Fa96} Falomo, R. 1996, MNRAS, 283, 241
\bibitem[Falomo \& Kotilainen (1999)]{FK99} Falomo, R., \& Kotilainen, J. 1999, A\&A, 352, 85
\bibitem[Falomo \& Ulrich (2000)]{FU00} Falomo, R., \& Ulrich, M.-H. 2000, A\&A, 357, 91
\bibitem[Fan et al. (2014)]{Fa14} Fan, J. H., Kurtanidze, O., Liu, Y., et al. 2014, ApJS, 213, 26

\bibitem[Feng et al. (2017)]{Fe17} Feng, H. C., Liu, H. T., Bai, J. M., et al. 2017, ApJ, 849, 161 

\bibitem[Fiorucci \& Tosti (1996)]{FT96} Fiorucci, M., \& Tosti, G. 1996, A\&AS, 116, 403

\bibitem[Fossati et tal. (1998)]{Fo98} Fossati, G., Maraschi, L., Celotti, A., et al. 1998, MNRAS, 299, 433

\bibitem[Ghisellini et al. (1998)]{Gh98} Ghisellini, G., Celotti, A., Fossati, G., et al. 1998, MNRAS, 301, 451
\bibitem[Gupta et al. (2008a)]{Gu08a} Gupta, A. C., Fan, J. H., Bai, J. M., et al. 2008a, ApJ, 135, 1384
\bibitem[Gupta et al. (2008b)]{Gu08b} Gupta, A. C., Deng, W. G., Joshi, U. C., Bai, J. M., \& Lee, M. G. 2008b, NewA, 13, 375
\bibitem[Gupta et al. (2012)]{Gu12} Gupta, S. P., Pandey, U.S., Singh, K., et al, 2012, NewA, 17, 8
\bibitem[Heidt et al. (1999)]{He99} Heidt, J., Nilsson, K., Sillanp\"{a}\"{a}, A., et al. 1999, A\&A, 341, 683
\bibitem[Howell (1989)]{Ho89} Howell, S. B. 1989, PASP, 101, 616
\bibitem[Hyv\"{o}nen et al. (2007)]{Hy07} Hyv\"{o}nen, T., Kotilainen, J. K., Falomo, R., et al. 2007, A\&A, 476, 723
\bibitem[Konopelko et al. (2003)]{Ko03} Konopelko, A., Mastichiadis, A., \& Kirk, J. 2003, ApJ, 597, 851
\bibitem[Kotilainen \& Falomo (2004)]{KF04} Kotilainen, J., \& Falomo, R. 2004, A\&A, 424, 107
\bibitem[Liu \& Bai (2015)]{Li15} Liu, H. T., \& Bai, J. M., 2015, AJ, 149, 191
\bibitem[Makino et al. (1990)]{Ma90} Makino, J., Akiyama, K., \& Sugimoto, D. 1990, PASJ, 42, 205

\bibitem[Maraschi \& Tavecchio (2003)]{Ma03} Maraschi, L., \& Tavecchio, F., 2003, ApJ, 593, 667

\bibitem[Neronov et al. (2012)]{Ne12} Neronov, A., Semikoz, D., \& Taylor, A. M. 2012, A\&A, 541, 31
\bibitem[Nilsson et al. (1999)]{Ni99} Nilsson, K., Pursimo, T., Takalo, L. O., et al. 1999, PASP, 111, 1223
\bibitem[Nilsson et al. (2003)]{Ni03} Nilsson, K., Pursimo, T., Heidt, J., et al. 2003, A\&A, 400, 95
\bibitem[Nilsson et al. (2007)]{Ni07} Nilsson, K., Pasanen, M., Takalo, L. O., et al. 2007, A\&A, 475, 199
\bibitem[Quinn et al. (1996)]{Qu96} Quinn, J., Akerlof, C. W., Biller, S., et al. 1996, ApJL, 456, L83
\bibitem[Samuelson et al. (1998)]{Sa98} Samuelson, F. W., Biller, S. D., Bond, I. H., et al. 1998, ApJ, 501, L17
\bibitem[Scarpa et al. (2000)]{Sc00} Scarpa, R., Urry, C., Padovani, P., Calzetti, D., \& O'Dowd, M. 2000, ApJ, 544, 258
\bibitem[S\'{e}rsic (1968)]{Se68} S\'{e}rsic, J.-L. 1968, Atlas de Galaxias Australes (Cordoba: Obs. Astron.)
\bibitem[Shukla et al. (2015)]{Sh15} Shukla, A., Chitnis, V. R., Singh, B. B., et al. 2015, ApJ, 798, 2
\bibitem[Stickel et al. (1993)]{St93} Stickel, M., Fried, J. W., \& K\"{u}ehr, H. 1993, A\&AS, 98, 393
\bibitem[Urry \& Padovani (1995)]{UP95} Urry, C. M., \& Padovani, P. 1995, PASP, 107, 803
\bibitem[Urry et al. (1998)]{Ur98} Urry, C. M., Falomo, R., Scarpa, R., et al. 1998, ApJ 512, 88
\bibitem[Urry et al. (2000)]{Ur00} Urry, C. M., Scarpa, R., O'Dowd, M., et al. 2000, ApJ, 532, 816
\bibitem[Villata et al. (1998)]{Vi98} Villata, M., Raiteri, C. M., Lanteri, L., et al. 1998, A\&AS, 130, 305
\bibitem[Xie et al. (1999)]{Xi99} Xie, G. Z., Li, K. H., Zhang, X., Bai, J. M., \& Liu, W. W. 1999, ApJ, 522, 846
\bibitem[Xie et al. (2001)]{Xi01} Xie, G. Z., Li, K. H., Bai, J. M., et al. 2001, ApJ, 548, 200
\bibitem[Xiong et al. (2016)]{Xi16} Xiong, D. R., Zhang, H. J., Zhang, X., et al. 2016, ApJS, 222, 24
\bibitem[Zhang et al. (2004)]{Zh04} Zhang, X., Zhang, L., Zhao, G., et al. 2008, AJ, 128, 1929
\bibitem[Zhang et al. (2008)]{Zh08} Zhang, X., Zheng, Y. G., Zhang, H. J., \& Hu, S. M. 2008, ApJS, 174, 111

\end{thebibliography}
\end{document}